\documentclass[twocolumn,tighten,times]{aastex62}

\usepackage{graphicx}
\usepackage{amssymb,amsfonts,amsmath}
\usepackage{float}
\usepackage{enumitem}
\usepackage{verbatim}
\usepackage{pifont}
\newcommand{\xmark}{\ding{55}}

\journalinfo{This Preprint originally submitted to AAS Journals, May 25, 2018 - See published version for peer-reviewed Version of Record.}

\shorttitle{Neutral Helium Spectroscopy of Coronal Rain}
\shortauthors{Schad}
\graphicspath{ {./} }

\begin{document}
\title{\large \textbf{Neutral Helium Triplet Spectroscopy of Quiescent Coronal Rain \\ with Sensitivity Estimates for Spectropolarimetric Magnetic Field Diagnostics}}

\correspondingauthor{Thomas Schad}
\email{schad@nso.edu}

\author[0000-0002-7451-9804]{Thomas A. Schad}
\affil{National Solar Observatory, 8 Kiopa`a Street, Ste 201, Pukalani, HI 96768, USA}

%%%%%%%%%%%%%%%%%%%%%%%%%%%%%%%%%%%%%%%%%%%%%%%%%%%%%%%%%%%%%%%%%%%

\begin{abstract}
On account of its polarizability and magnetic field sensitivity, as well as the role of neutral helium in partially ionized solar environments, the neutral helium triplet (orthohelium) system provides important, yet under-utilized, diagnostics of solar coronal rain.  This work describes off-limb observations of coronal rain in NOAA Active Region 12468 obtained in the \ion{He}{1} 10830\mbox{\AA} triplet using the \textit{Massively MultipleXed Imaging Spectrograph} (MXIS) experiment at the Dunn Solar Telescope along with co-temporal observations from NASA's \textit{Solar Dynamics Observatory} (SDO) and the \textit{Interface Region Imaging Spectrograph} (IRIS).  We detect rain simultaneously in the IRIS 1400\mbox{\AA} and 2796\mbox{\AA} channels and in \ion{He}{1} 10830\mbox{\AA}.  The large degree of spatial coherence present between all channels agrees with previous observations of the multi-thermal nature of coronal rain.  A statistical analysis of \ion{He}{1} spectral profiles for rain identified via automated detection indicate \ion{He}{1} line radiances are, on average, $10^{4}$ ergs cm$^{-2}$ s$^{-1}$ sr$^{-1}$; the average translational velocity is 70 km s$^{-1}$ and Doppler widths are distributed around 10 km s$^{-1}$.  Based on these results, forward models of expected \ion{He}{1} polarized signals allow us to estimate, using synthetic observables and an inversion algorithm including fits for the scattering angle constraining the material's location along-the-line of sight, the magnetic sensitivity of the upcoming National Science Foundation's Daniel K. Inouye Solar Telescope (DKIST).  We predict joint observations of the \ion{He}{1} 10830\mbox{\AA} and 5876\mbox{\AA} multiplets, using first-light instrumentation, will provide inverted magnetic field errors of $\pm3.5$ G ($2\sigma$) for spatial scales of $0\farcs5$ ($\sim360$ km) assuming dynamically-limited integration times of 5.5 seconds.
\end{abstract}


%%%%%%%%%%%%%%%%%%%%%%%%%%%%%%%%%%%%%%%%%%%%%%%%%%%%%%%%%%%%%%%%%%%

\section{Introduction}

Remote sensing the megakelvin solar corona is difficult as the corona is primarily optically thin and the responsiveness of available diagnostics, especially for the magnetic field dominating its evolution, is limited. This impedes our ability to answer fundamental questions regarding how stellar coronae are structured and energized.  Remarkably, the production of cool material in the otherwise hot corona gives us perhaps our finest probe of its structure.  Observations have shown with increasing detail that the $>$1MK active corona frequently produces condensed, cool (10s of kK$)$, gravitationally-unstable material known as coronal rain \citep{kawaguchi1970, schrijver2001, antolin2012sharp}.  Thought to be caused by a thermal instability \citep{field1965,goldsmith1971,antiochos1991,schrijver2001}, it can be formed under multiple scenarios with major categories including quiescent (or non-eruptive) and flare-driven, which is typically stronger (brighter) \citep{scullion2016}.  Its formation is also sometimes associated with prominence/filament dynamics \citep{liu2012}, for example, in the draining spider `legs' of coronal cloud filaments \citep{schad2016}. Often, though, it materializes directly from the hot corona.  During these fast ($\sim$10-150 km s$^{-1}$) rain events, coronal material is both localized and bright in cool, radiatively-excited chromospheric spectral lines, which in turn provide a valuable probe of local conditions during non-equilibrium.  

Helium, as the second most abundant element, generates relatively bright emission from coronal rain.  It is chiefly observed using EUV imaging of \ion{He}{2} 304\mbox{\AA} ($\log T \approx 4.7$ K) \citep{degroof2004,degroof2005,kamio2011}.  In comparison, neutral helium coronal rain emission has not been studied in detail.  Neutral material observed in coronal rain blobs as they fall, both on and off-disk, has been best quantified in H$\alpha$ \citep{antolin2012sharp, antolin2012ondisk}, which generally exhibits dynamic coupling with the ionized species of \ion{He}{2} 304\mbox{\AA} \citep{degroof2005}, \ion{Ca}{2} 8542\mbox{\AA} \citep{ahn2014}, and \ion{Ca}{2} H \citep{chae2010}, in addition to \ion{C}{2}, \ion{Si}{4}, and \ion{Mg}{2}, which dominate the IRIS 1330\mbox{\AA}, 1400\mbox{\AA}, and 2796\mbox{\AA} imaging filters \citep{antolin2015}.  Sizeable fractions of neutrals can potentially affect the plasma dynamics, and models by \cite{zaqarashvili2011} indicate the formation of neutral helium in particular can enhance the damping of Alfv{\'e}n waves in partially ionized plasmas.  Transverse oscillations observed within coronal rain have been studied by \cite{antolin2011}, \cite{okamoto2015}, and \cite{kohutova2017}.

The resonance lines of neutral helium are formed in the EUV below 584\mbox{\AA} and are not currently observed routinely \citep{doschek1974, peter1999a,judge2004}; however, for characteristic densities of condensed coronal material, \textit{i.e.} prominences or coronal rain, photo-ionization followed by recombination into the triplet system, whose lowest level ($2s^{3}S$) is metastable, overpopulates the triplet in comparison to singlet helium \citep{heasley1974, andretta1997, labrosse2001, gouttebroze2009}.  As a result, the 10830\mbox{\AA} multiplet is the brightest neutral helium feature, therefore making it a promising candidate for studying neutral helium production in coronal rain.  A further advantage is that is can be accessed using ground-based facilities.  

Due to its brightness and polarizability, the \ion{He}{1} triplet may also potentially be used to remote sense the magnetic field of coronal condensates.  \cite{antolin2012sharp} and \cite{scullion2016} both call attention to the substructure of coronal loops revealed by coronal rain and unseen in hot EUV observations, thereby designating rain an instrument for fine-scaled coronal magnetism studies.  The helium triplet is routinely employed to probe chromospheric magnetic fields \citep{harvey1971, asensio_ramos2008, schad2013} including within prominences \citep{casini2003}.  Its polarized signatures result from the joint action of resonant optical pumping, the Zeeman effect, and Hanle effect, and offers sensitivity for the vector field for a wide range of field strengths.  In the case of \emph{quiescent} coronal rain, short dynamical timescales and weak signals drive the need for large-aperture polarimetry even for the brightest lines to achieve a signal-to-noise sufficient to measure the weak field strengths of the solar corona.  The large collecting area of the National Science Foundation's 4 meter aperture Daniel K. Inouye Solar Telescope  \citep[DKIST:][]{rimmele2015}, currently under construction, is poised to greatly enhance such capabilities. 

Owing in large part to the use of novel multiplexed spectroscopic techniques \citep{schad2017_inst}, the study presented here performs time-resolved spectroscopy of \ion{He}{1} 10830\mbox{\AA} observed in \emph{quiescent} coronal rain off-limb.  This permits analysis of its evolving morphology as well as statistical quantification of its total integrated line radiance (brightness),  spectral line width, and both its apparent and Doppler velocity characteristics.  Coordinated IRIS observations are used to compare the neutral helium characteristics with those observed in ionized species of multiple temperatures.  Based upon the \ion{He}{1} results, inversion techniques and an error model are developed to assess the feasibility of using polarized He {\small \sc I} measurements of quiescent coronal rain off limb to infer the coronal magnetic field intensity and orientation, as well as the location of the coronal rain blobs along the line-of-sight.

%%%%%%%%%%%%%%%%%%%%%%%%%%%%%%%%%%%%%%%%%%%%%%%%%%%%%%%%%%%
%%%%%%%%%%%%%%%%%%%%%%%%%%%%%%%%%%%%%%%%%%%%%%%%%%%%%%%%%%%
%%%%%%%%%%%%%%%%%%%%%%%%%%%%%%%%%%%%%%%%%%%%%%%%%%%%%%%%%%%

\begin{figure*}[htb!]
\centering
\includegraphics[width=0.975\textwidth]{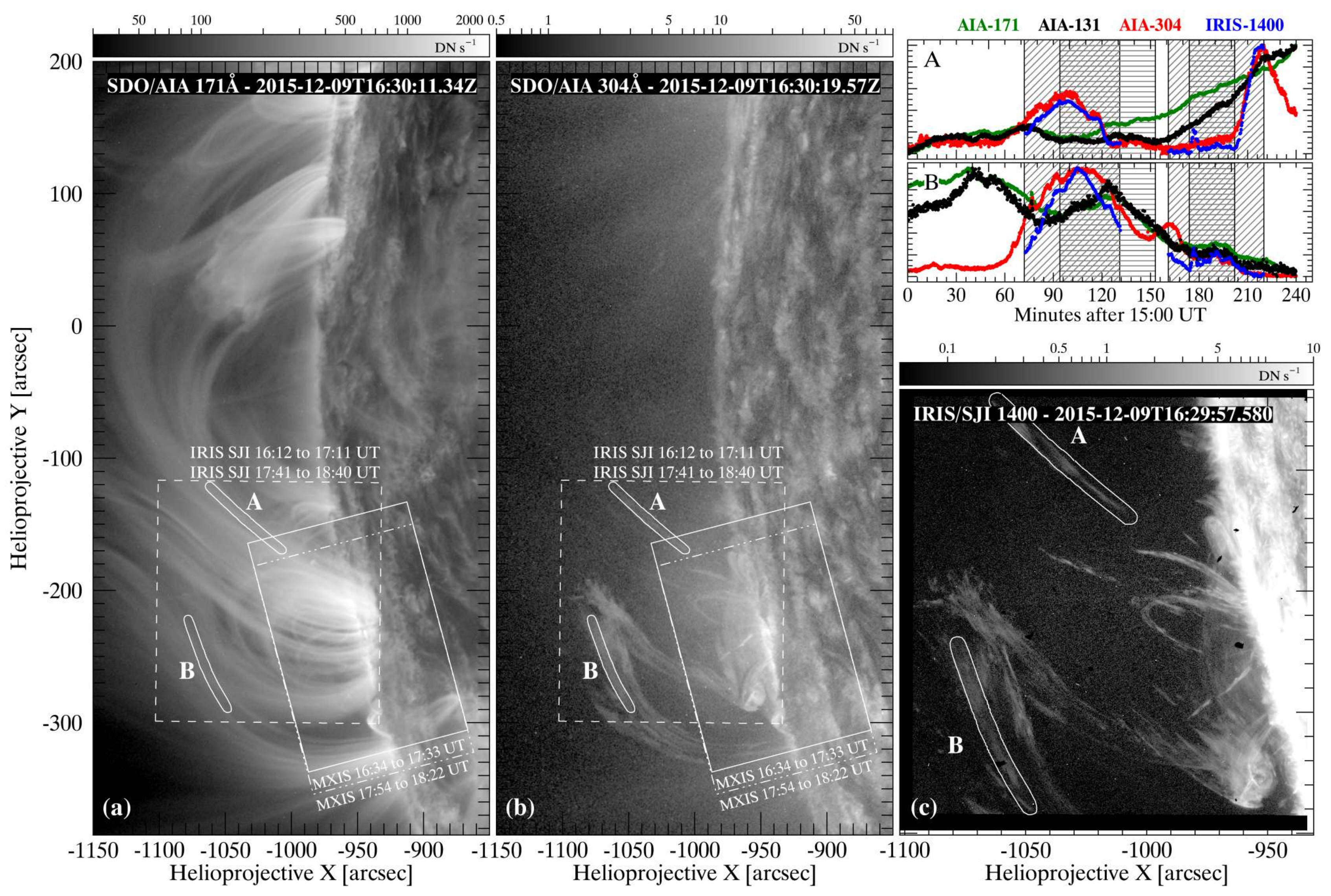}
\caption{Snapshots of coordinated observations targeting coronal rain on 2015 December 9 near 16:30 UT.  (a) AIA 171\mbox{\AA} ($\log T\approx 5.8$) images show hot coronal loops above NOAA AR 12468 in the southern hemisphere as well as trans-equatorial loops connected to plage in the northern hemisphere.  (b) AIA 304\mbox{\AA} ($\log T\approx 4.7$) images display coronal rain being produced in this region along with an active region prominence just off-limb. (c) IRIS SJI 1400\mbox{\AA} observations.  Regions of interest `A' and `B' demark sample locations of coronal rain evolution present in both AIA and IRIS; light curves averaged over these regions are shown in the upper right panel.  \\
(An animation of this figure is available)}
\label{fig:obs_fovs}
\end{figure*}

\section{Observations}

High-resolution observations of coronal rain were obtained on 2015 December 9 within the \ion{He}{1} 10830\mbox{\AA} spectral lines by the \textit{Massively MultipleXed Imaging Spectrograph} (MXIS), an experimental infrared imaging spectrograph at the 76 cm aperture Dunn Solar Telescope (DST) in New Mexico, USA.  The MXIS field-of-view was centered on the eastern solar limb and NOAA Active Region 12468.  These observations were coordinated with the \textit{Interface Region Imaging Spectrograph (IRIS)} \citep{depontieu2014}, and also make use of the high-cadence imaging data provided by the Atmospheric Imaging Assembly \citep[AIA:][]{lemen2011} and Helioseismic and Magnetic Imager  \citep[HMI:][]{schou2012} instruments onboard NASA's Solar Dynamics Observatory \citep[SDO:][]{pesnell2012}.  Figure~\ref{fig:obs_fovs} presents an overview of the fields-of-view and relative timing of the coordinated observations.

\subsection{Massively MultipleXed Imaging Spectrograph (MXIS)}\label{sec:he_obs}

The MXIS experiment, described in detail by \cite{schad2017_inst}, provides rapid-cadence, wide-field imaging spectroscopy of \ion{He}{1} 10830\mbox{\AA} by taking advantage of multiplexing techniques developed for full-disk spectroheliography \citep{lin2014}.  It consists of two science channels---a grating-based spectrograph and a narrowband imager\textemdash simultaneously imaged on separate halves of a single $2048\times2048$ pixel HgCdTe detector.  The detector is operated at a frame rate of 9.53 Hz with exposure times of 105 msec.

The MXIS spectrograph channel utilizes 17 parallel entrance slits separated by $11\farcs4$ over which a $175\arcsec  \times 125\arcsec$ field-of-view is scanned by a field steering mirror.  The great number of slits allows the large field-of-view to be mapped at high spatial resolution using only 65 discrete steps at a fast cadence ($\approx$8.5 seconds per full field scan).  Each slit has a full-width half-maximum spectral bandwidth of 10.5\mbox{\AA} allowing coverage of the \ion{Si}{1} photospheric line at 10827\mbox{\AA} and the \ion{He}{1} triplet.  The \ion{He}{1} Doppler coverage is -195 to 85 km s$^{-1}$.  The spectral pixel width is 120.4 m\mbox{\AA}; though, spectral resolution is limited for the data described here by spectral focus aberrations suffered during this campaign, as discussed by \cite{schad2017_inst} and carefully calibrated in Section~\ref{sec:lsf}.  The slit width is $\approx0\farcs19$ while the angular sampling along the slit is $0\farcs153$. 

The MXIS narrowband imaging channel operates using about $1\%$ of the light fed to MXIS, which is reflected by an optical wedge upstream of the spectrograph.  With a set of independent reimaging optics, the spatial scale incident on the detector is $0\farcs123$ pixel$^{-1}$. The spectral bandpass is identical to the spectrograph as the filters are located upstream of both channels.  Since that bandwidth is dispersed over $\approx90$ detector pixels in the spectrograph, the raw signal to noise in the two channels is approximately balanced.

Two data series, spanning 16:34 - 17:33 UT and 17:54 to 18:22 UT, were obtained by MXIS on 2015 December 9 overlaping a portion of the IRIS field-of-view (see Figure~\ref{fig:obs_fovs}).  Active tip/tilt correction for MXIS necessitated the spatial offset of the field relative to IRIS since the DST High Order Adaptive Optics (HOAO) system \citep{rimmele2004} requires the limb to be positioned near the field center for active tracking.  Data reduction followed the techniques outlined in \cite{schad2017_inst} including precise co-registration between the two channels and the post-facto correction of residual seeing-induced image motion.  However, for these off-limb data, only a scalar correction for tip/tilt variability on a frame by frame basis is applied;  sub-field seeing-induced motion is not corrected. 

\begin{figure}[htb!]
\centering
\includegraphics[width=0.475\textwidth]{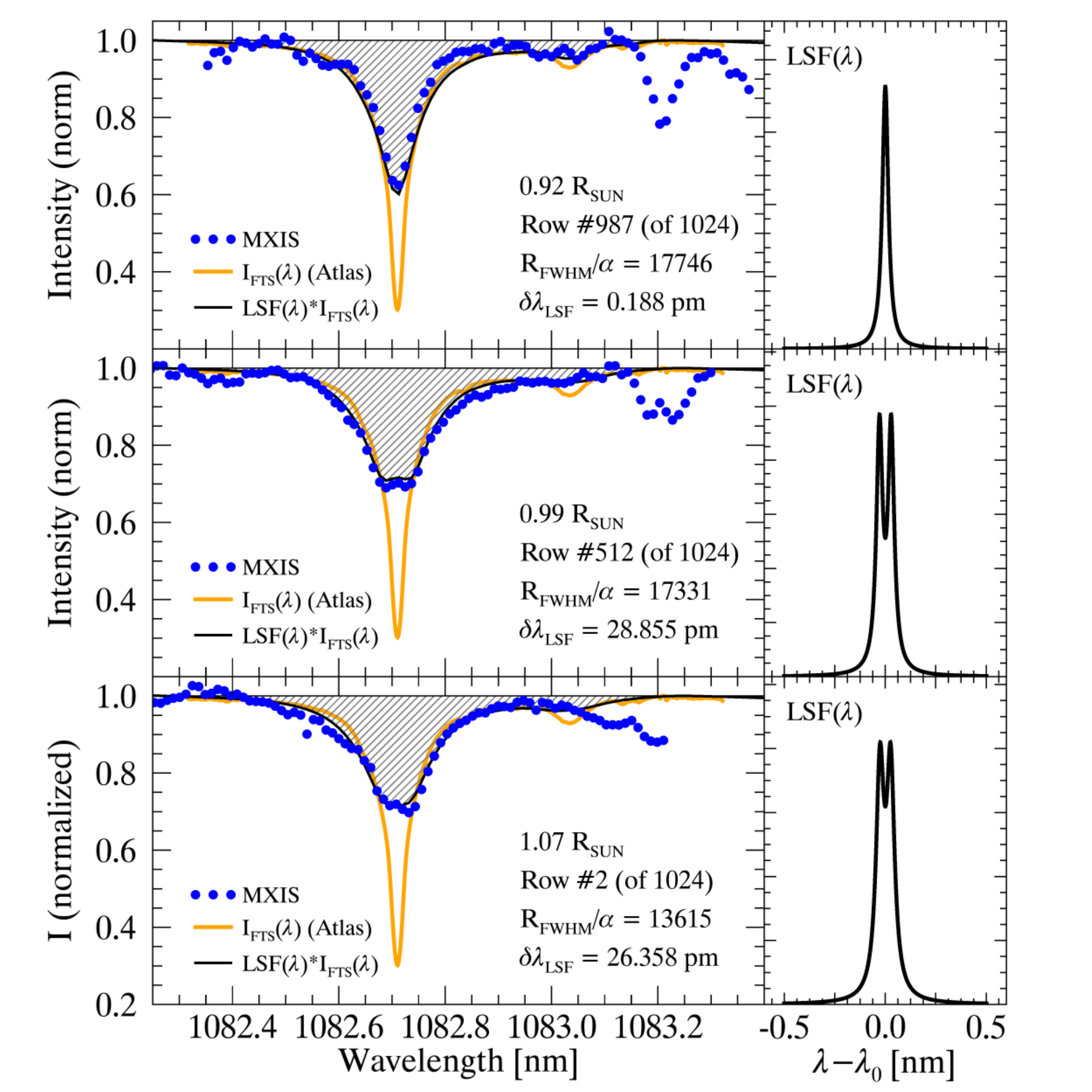} 
\caption{MXIS spectrograph line spread function (LSF) calibration.  Each row shows a different quiet-sun disk center profile extracted along the center spectrograph slit, its derived LSF, and the convolved atlas spectrum.  The location of these profiles relative to the solar limb for the coronal rain observations is also indicated.  $R_{\rm FWHM}$ is the ratio of wavelength to full-width half maximum of the Lorentzian functions used in the LSF.  $\alpha=1.7$, as in \cite{schad2017_inst}, and $\delta\lambda_{LSF}$ is the separation factor approximating the spectrograph defocus.}
\label{fig:spectral_calib}
\end{figure}

\subsubsection{Line spread function calibration}\label{sec:lsf}

MXIS spectrograph data acquired in December 2015 was affected by a suboptimal spectrograph focus limiting the spectral resolution ($R$) to between 8,000 and 18,000, considerably lower than subsequently acquired data ($R$$\approx$25,000).  While these aberrations do not affect the measured integrated line radiances, to facilitate the fitting of models to the observed profiles and the measurement of true spectral line widths, we have determined a suitable line spread function (LSF) describing the spectrograph response, which varies across the field of view.  Examples of the LSF calibration are shown in Figure~\ref{fig:spectral_calib}.  Following \cite{schad2017_inst}, the spectrograph LSF can be approximated by a Lorentzian function; however, to account for the defocus, we instead consider here an LSF modeled by two equal Lorentzian functions separated in wavelength by a small amount.  This separation factor approximates the removal of energy from the core of the LSF due to the defocus.  Using this functional form, we optimize the parameters of the LSF by fitting, in a least-squares fashion, an LSF-convolved high resolution solar atlas spectrum of the deep \ion{Si}{1} 10827\mbox{\AA} line to the measured spectral profiles obtained during flat-field observations near disk center.  The figure shows the resulting LSF and the fits for three spatial locations along the center MXIS slit.  The corresponding radial locations in the science data described here are noted in the figure.  All modeled fits to spectral profiles discussed below include convolution with the field-dependent LSF.

\begin{figure}
\centering
\includegraphics[width=0.475\textwidth]{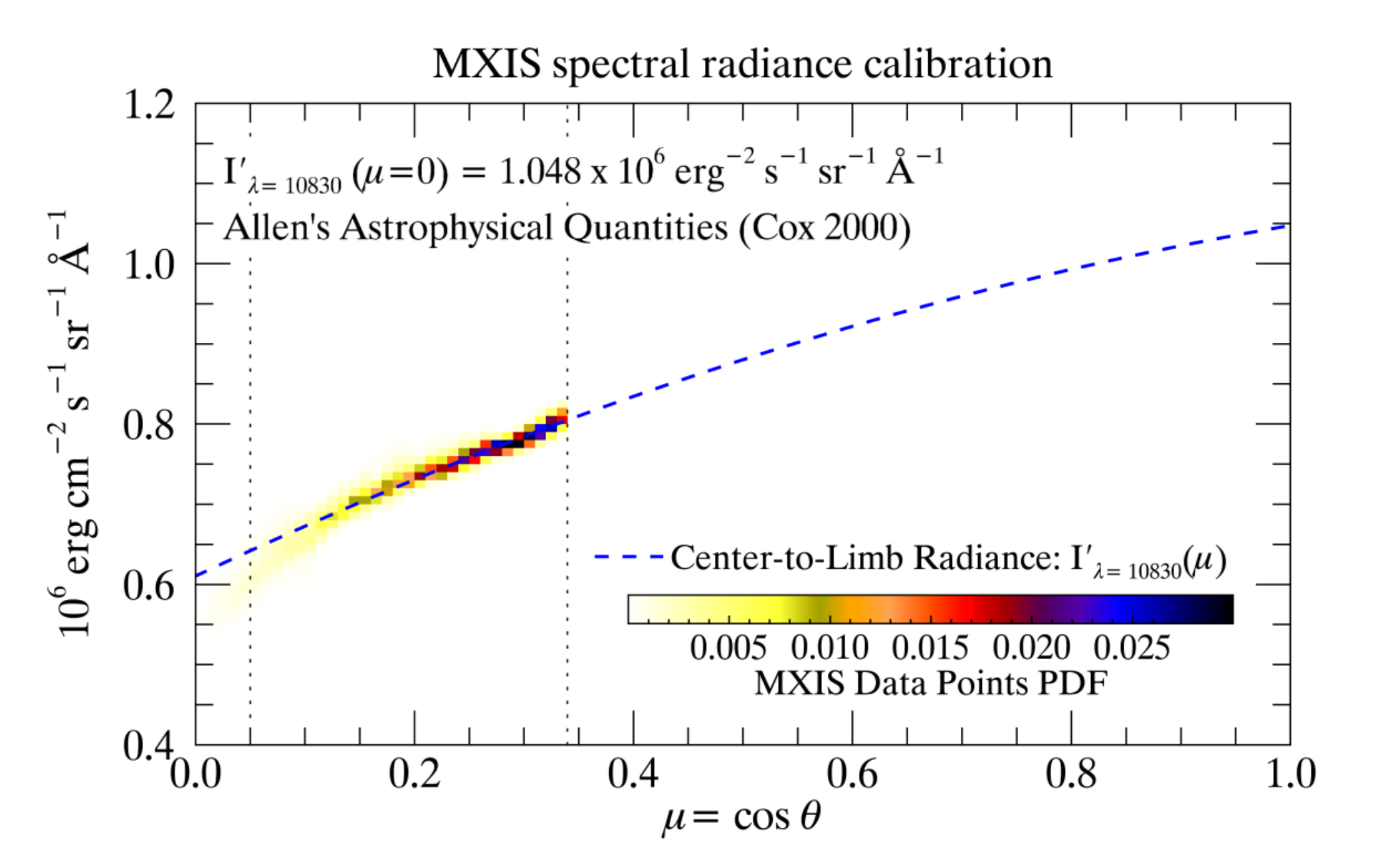}
\caption{Radiometric calibration of MXIS \ion{He}{1} spectrograph data values showing the two-dimensional probability distribution function of continuum values for all on-disk data points within a single MXIS field scan and their relationship to the known center-to-limb dependence of the spectral radiance in the continuum near 10830\mbox{\AA}.} 
\label{fig:photo_calib}
\end{figure}

\subsubsection{Radiometric calibration}

To convert observed flux from reduced MXIS spectrograph data units to units of absolute spectral radiance, we derive a calibration factor by comparing the center-to-limb flux variation measured in the solar continuum for $0.05 < \mu <0.3$, where $\mu$ is the cosine of the heliocentric angle, with the well known center-to-limb variation of the solar continuum spectral intensity at 10830\mbox{\AA}, as available in Allen's Astrophysical Quantities \citep{cox2002}.  Since we observe a large portion of the solar limb during every scan, this calibration is done on a scan-by-scan basis and thus eliminates any dependencies on temporal sky transmission variability.  An example of the converted data units in the continuum and the comparison to the center-to-limb reference curve is given in Figure~\ref{fig:photo_calib}.

\subsection{SDO/AIA and SDO/HMI}

The SDO level 1.5 data products from AIA and HMI are used to establish the global evolutionary context of the \ion{He}{1} 10830\mbox{\AA} MXIS observations as well as for image registration steps described in Section~\ref{sec:data_coalign}.  We focus mainly on the AIA 171\mbox{\AA}, 131\mbox{\AA}, and 304\mbox{\AA} EUV channels and use the 1700\mbox{\AA} UV channel to facilitate alignment with IRIS.  For non-flaring conditions, the 171\mbox{\AA}, 131\mbox{\AA}, and 304\mbox{\AA}  AIA passbands are dominated by \ion{Fe}{9}, \ion{Fe}{8}, and \ion{He}{2} emission, with respective characteristic temperatures of $\log T  = 5.8$, $5.6$, and $4.7$ K \citep{odwyer2010}.  The spatial scale of all level 1.5 data is $0\farcs6$ pixel$^{-1}$, and the cadence is 12 (24) seconds for EUV (UV) data, while the HMI intensitygrams have a cadence of 45 seconds.  Near the observed region, the field-of-view of AIA extends to approximately 180 Mm above the limb, \textit{i.e.} $\gtrsim 1.26$ solar radii.  

\subsection{IRIS}

IRIS, which consists of both a spectrograph (SG) and a slit-jaw imager (SJI), conducted two observation series, spanning 16:12 - 17:11 UT and 17:41 - 18:40 UT, with its pointing centered at $[x,y] = [-1017\arcsec,-209\arcsec]$.  Both series used the OBS 3620259404 program operating both the SG and SJI; though, here we only use the level 2 SJI data consisting of a very large ($167\arcsec \times 174\arcsec$) sit-and-stare image sequence cycling between the SJI 1400\mbox{\AA} and 2796\mbox{\AA} filters.  The 55\mbox{\AA} wide passband of the 1400\mbox{\AA}  filter is dominated by the transition region \ion{Si}{4} spectral lines at 1393.78\mbox{\AA} and 1402.77\mbox{\AA}, which are formed near $\log T = 4.8$, while the 2796\mbox{\AA}  filter has a 4\mbox{\AA} passband dominated by the \ion{Mg}{2} k line core at 2796.35\mbox{\AA} formed near $\log T = 4$ \citep{depontieu2014}.  The temporal cadence for each filter is 19 seconds, and the image scale is $0\farcs166$ pixel$^{-1}$. Spatial resolution is limited to $0\farcs33$ for the 1400\mbox{\AA}  channel and $0\farcs4$ for the 2796\mbox{\AA}  channel. Quick-look summary information can be accessed via the Heliophysics Events Knowledgebase Coverage Registry(HCR)\footnote{Follow these links to reach the HCR provided IRIS observation details: \url{https://tinyurl.com/iris-rain-series1}; \url{https://tinyurl.com/iris-rain-series2}}.  

\subsection{Data co-alignment}\label{sec:data_coalign}

Pointing instability limits the accuracy of the image coordinates relative to a fixed solar frame for both IRIS and MXIS.  These IRIS data were acquired during eclipse season when IRIS slit jaw images exhibit amplified pointing wobble.  For correction, the AIA 1700\mbox{\AA} image nearest in time to each frame of the IRIS 1400\mbox{\AA}  channel was remapped into the spatial header coordinates of the first frame in each fixed-pointing IRIS 1400\mbox{\AA}  data series.  A scalar shift is then determined using cross-correlation, as recommended in IRIS Technical Note 22\footnote{\url{https://www.lmsal.com/iris_science/doc?cmd=dcur&proj_num=IS0212&file_type=pdf}}, that best aligns high-pass filtered versions of the IRIS 1400\mbox{\AA} image and the AIA 1700\mbox{\AA} image. The spatial offset of each frame of the 2796\mbox{\AA}  channel is approximated by that of the nearest 1400\mbox{\AA}  frame.  For correction, the spatial offsets are applied to the level 2 data. 

In the MXIS data, the HOAO active tip/tilt stabilization offered fair correction in the direction perpendicular to the limb, but drifts and seeing-induced jumps in the direction parallel to the limb remain.  We correct the pointing perturbations by cross-correlating HMI intensitygrams with each MXIS narrowband context image.  A general polynomial warping transformation is first derived between a single HMI intensitygram and a co-temporal MXIS narrowband context image using the SSWIDL routine \textit{auto\_align\_images}.  The images were selected based on the period of best seeing during the MXIS data series.  Using this transformation, an HMI intensitygram is remapped into the MXIS coordinate frame for each time-step of the MXIS data series.  Then, a scalar shift between the remapped HMI intensitygram and the MXIS context image is derived via cross-correlation and used to correct pointing errors in the MXIS data.  This procedure overcomes the difficulty of self-aligning the MXIS time series that arises due to the evolution of the off-limb structures and solar rotation. 

%%%%%%%%%%%%%%%%%%%%%%%%%%%%%%%%%%%%%%%%%%%%%%%%%%%%%%%%%%%%%%%%%%%%
%%%%%%%%%%%%%%%%%%%%%%%%%%%%%%%%%%%%%%%%%%%%%%%%%%%%%%%%%%%%%%%%%%%%
%%%%%%%%%%%%%%%%%%%%%%%%%%%%%%%%%%%%%%%%%%%%%%%%%%%%%%%%%%%%%%%%%%%%

\begin{figure*}[htb!]
\centering
\includegraphics[width=0.975\textwidth]{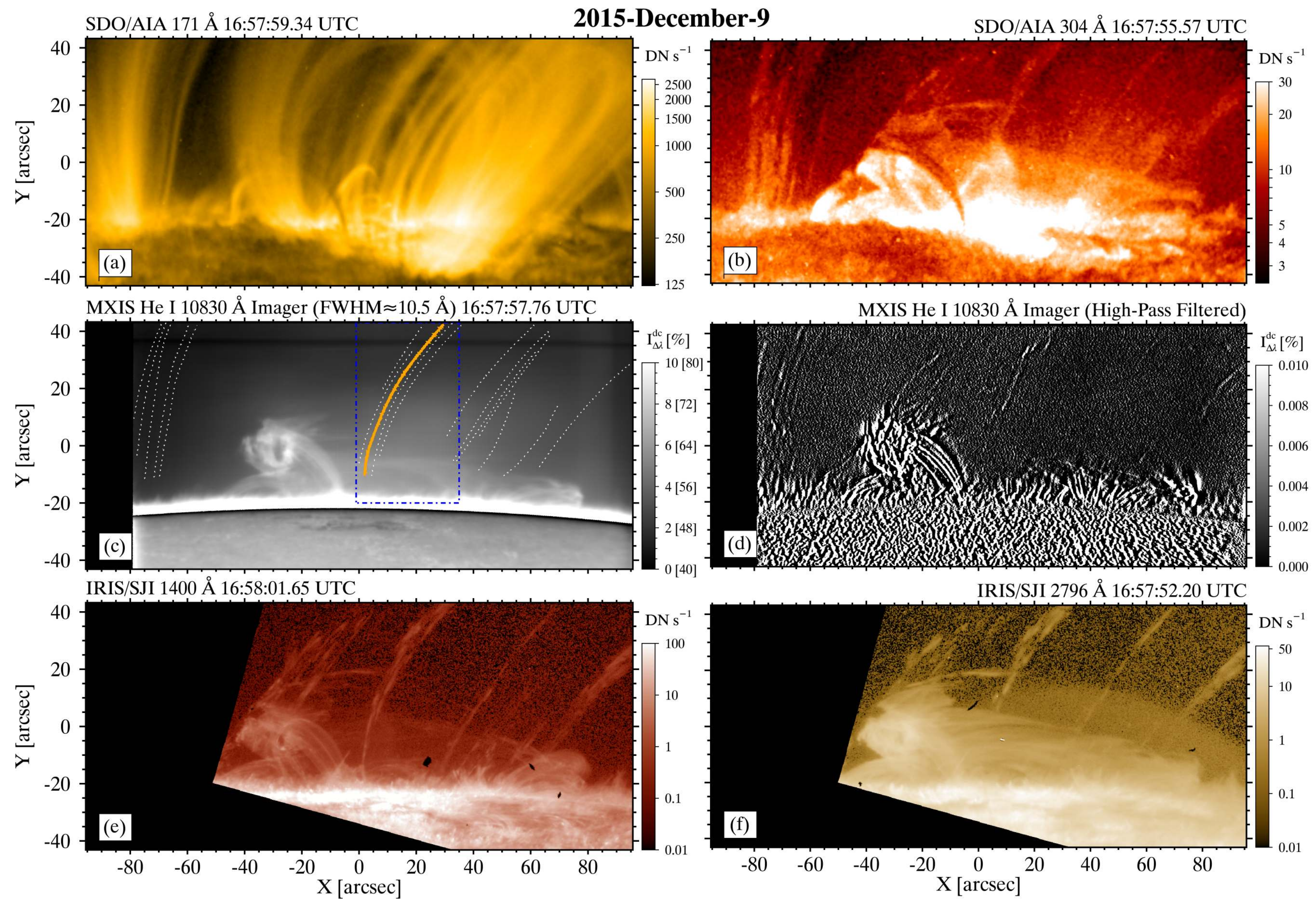}
\caption{Multi-wavelength comparison of the observed coronal rain demonstrating the correspondence of the neutral helium production along the same trajectories as the ultraviolet diagnostics of ionized species.  Snapshots from the associated animation near 16:58 UT are shown for (a) AIA 171\mbox{\AA}, (b) AIA 304\mbox{\AA}, (c) the MXIS narrowband imaging channel covering \ion{He}{1} 10830\mbox{\AA}, (d) a high-passed filtered version of the same, (e) IRIS SJI 1400\mbox{\AA}, and (f) IRIS SJI 2796\mbox{\AA}.  All intensity scale values for MXIS are given as a percentage of $I_{\Delta\lambda}^{dc}$, the total intensity integrated over the imager bandwidth at disk center.  Dotted lines in (c) denote manually traced loops representative of downflows of \ion{He}{1} rain.  The loop drawn as a solid orange line (within the blue dot-dashed sub-region) is examined further in Figures~\ref{fig:time_series} and~\ref{fig:time_slices}.\\ 
(An animation of this figure is available)}
\label{fig:all_data_aligned}
\end{figure*}

\section{Temporal Evolution and Rain Formation in NOAA AR 12468}

In the AIA 171\mbox{\AA} data (see Figure~\ref{fig:obs_fovs} and its associated animation) NOAA AR 12468 displays bright loops with transequatorial linkage to distributed plage in the northern hemisphere; the loops exhibit similarities to rain-producing loops studied by \cite{auchere2018}.  The presence of coronal rain is easily discerned using \ion{He}{2} 304\mbox{\AA} observations \citep{degroof2004,degroof2005,kamio2011}, and the time series of AIA 304\mbox{\AA} observations between 16:00 UT and 18:40 UT reveal apparent raining down-flows in various parts of the active region including both foot-points of the transequatorial loops.  Rain originating from the transequatorial loops as well as from structures extending to the edge of the field of view directly above the active region is apparent in the AIA 131\mbox{\AA}, 171\mbox{\AA}, and 304\mbox{\AA} channels and in the IRIS SJI data (see, \textit{e.g.} regions A and B in the figure).    

At approximately 16:00 UT, a solar surge \citep[see, \textit{e.g.}][]{kirshner1971,roy1973} erupts from the southwestern edge of the IRIS field-of-view and extends outwards from the solar limb.  Components of the surge can be seen in all channels.  An apparent down-flow similar to the coronal rain material exists in the wake of the surge that persists until approximately 17:10 UT.  While this material is observed in the MXIS \ion{He}{1} observations as well, it is not included in our analysis of coronal rain, especially since it is brighter in relation to the rain unassociated with the surge.  An active region prominence can be seen near the limb in all channels.  A background portion of the prominence becomes activated around 18:00 UT; while the foreground component remains stable throughout the time-period. 

One phenomenon indicative of coronal rain produced via thermal instability is the progressive illumination of cooler EUV channels \citep{kjeldseth_moe_1998, schrijver2001, kamio2011}.  Light curves displayed in the upper right portion of Figure~\ref{fig:obs_fovs} show spatial averages over the regions labeled A and B versus time.  In particular, region B gives evidence for progressive cooling as the hot AIA 171\mbox{\AA}  and 131\mbox{\AA}  channels increase in brightness prior to increases in brightness of the AIA 304\mbox{\AA}  and IRIS 1400\mbox{\AA}  channels.  The increased brightness of AIA 304\mbox{\AA}  and IRIS 1400\mbox{\AA}  in Region A around 16:30 UT is not preceded by a brightening in the hotter channels; however, it is likely that the onset of this coronal condensation commences near the transequatorial loop apex outside of the IRIS field-of-view.

Figure~\ref{fig:all_data_aligned} (and its associated animation) zooms in on the region observed by MXIS in coordination with AIA and IRIS.  MXIS observes only the lower portion of extended coronal loops, and thus a majority of the coronal rain observed enters the field-of-view having already been formed.  To describe the observed features, we first concentrate on the narrowband imaging channel of MXIS.  An example image averaged over one field scan (i.e. sixty-five 105 msec exposures) is given in panel (c) of the figure with units of percent of $I_{\Delta\lambda}^{dc}$, \textit{i.e.} the total disk center intensity integrated over the imager filter bandwidth.  A different scaling is used above the limb to enhance the off-limb features.  On the solar disk the filter's response is dominated by the limb-darkened photospheric continuum and the Si I absorption at 10827\mbox{\AA}---sunspots and plage are evident.  Off limb only the \ion{He}{1} signal contributes to the intensity, and thus spicules and the active region prominence are easily detected with intensities up to 10\% of $I_{\Delta\lambda}^{dc}$.

\ion{He}{1} coronal rain present in panel (c) is not readily apparent due to its low contrast.  A high pass filtered version of the MXIS narrowband filter image is shown in panel (d) to highlight the raining material.  The high-pass filter is created in two steps to mitigate instrument related systematics that affect the horizontal and vertical directions of the image in different ways.  First, the original image is smoothed in the horizontal direction (parallel to `X' axis in Figure~\ref{fig:all_data_aligned}) using a boxcar averaging kernel of 10 pixels ($1\farcs 23$).  The smoothed image is subtracted from the original image to create an intermediate image.  This image is then smoothed vertically with a boxcar averaging kernel of 100 pixels ($12\farcs3$) and subtracted from the intermediate image to create the final high-pass filtered image.  This process suppresses horizontally or vertically oriented structures--\textit{e.g.} the horizontal portion of the prominence body is no longer visible in panel (d)--but does well to enhance the coronal rain material, which is now easily seen above the forest of spicules in panel (d).  The intensities of the rain in the narrowband image are on order of $0.01\%$ of $I_{\Delta\lambda}^{dc}$.

The animation associated with Figure~\ref{fig:all_data_aligned} provides the best visual illustration of the widespread presence of neutral helium and furthermore demonstrates that neutral helium forms along the same trajectories where rain is present in AIA 304\mbox{\AA} and the IRIS channels.  The snapshots of Figure~\ref{fig:all_data_aligned} also give evidence for this correspondence.  The primarily vertical structures observed in all cool channels above the limb, \textit{i.e.} those other than 171\mbox{\AA}, consist of apparent coronal rain downflows. Of one exception is the material near $[X,Y] = [-20\arcsec,20\arcsec]$, which is associated with the aftermath of the solar surge. 
 
\section{Multi-Spectral Properties of Individual Downflows}

For more detailed analysis, a representative sample of 17 raining loops is manually identified and traced using the MXIS narrowband \ion{He}{1} imaging data (see loops overplotted in Figure~\ref{fig:all_data_aligned}(c).  For each traced loop, we extracted a time slice diagram from each coordinating instrument.  Upon inspection of the time slice diagrams, we found a high level of correspondence between the IRIS channels and \ion{He}{1} 10830\mbox{\AA}.  For some events that correspondence is confused by multiple evolving loops near the same location or by brief periods of poor seeing at the DST.  In other cases, no overlap in time exists for MXIS and IRIS.  Below, we closely examine a raining feature representative of the large-scale correspondence and identified by the solid orange line in Figure~\ref{fig:all_data_aligned} panel (c). 

\begin{figure*}[htb!]
\centering
\includegraphics[width=0.65\textwidth]{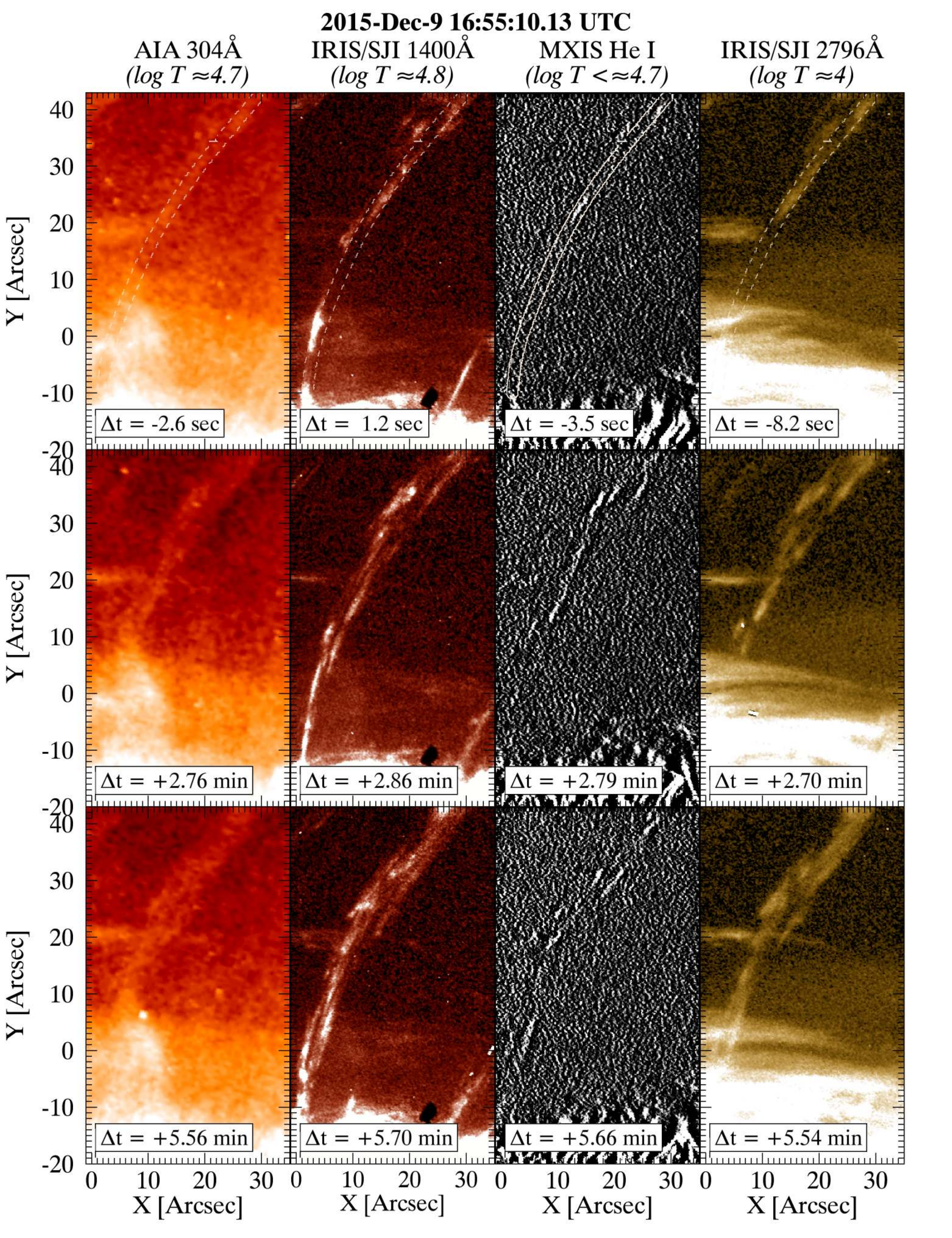}
\caption{Temporal evolution of the individual rain path identified by the solid orange line in Figure~\ref{fig:all_data_aligned}c) and for all observed cool channels.  For clarity, the orange line has been replaced by minimal dotted and solid outlines in the top row.  The MXIS \ion{He}{1} data corresponds to the narrowband imaging channel data after high-pass filtering as described in the text.  Each row is quasi-simultaneous. }
\label{fig:time_series}
\end{figure*}

\subsection{Morphological Comparison}

Figure~\ref{fig:time_series} provides snapshots of each cool channel within the $35\arcsec \times 60\arcsec$ ($25 \times  44$ Mm) blue dot-dashed sub-field of Figure~\ref{fig:all_data_aligned}(c) to illustrate the temporal evolution of the raining feature as a function of characteristic bandpass temperature.  This especially serves to put the neutral helium 10830\mbox{\AA} observations in context with the multi-thermal, finely-structure morphology of coronal rain discussed by \cite{antolin2015}.  The individual images in each row are separated by no more than 10 seconds in time, whereas each row is progressively separated by $\approx165$ seconds (2.75 min).  The translational apparent speeds of the material are between 40 and 95 km s$^{-1}$ [$0\farcs05-0\farcs13$ s$^{-1}$] (derived below in Section~\ref{sec:time_slices}), and therefore these images follow the evolution of multiple rain blobs as they fall in what is referred to as a rain shower. 

Morphologically the coronal rain observed in \ion{He}{1} 10830\mbox{\AA} shares many of the characteristics shown within the IRIS SJI 1400\mbox{\AA}  and 2796\mbox{\AA}  data including a significant degree of spatial coherence.  See, among other examples, the bright kernels in the middle row of Figure~\ref{fig:time_series} near $[X,Y]=[18\arcsec,34\arcsec]$.  Here, we note that under conditions of collisional ionization equilibrium \citep[see, \textit{e.g.} tables in Chianti v.8,][]{delzanna2015}, the ionization fraction of \ion{Si}{4} only weakly overlaps with that neutral helium and at fractions less than 3\% of the peak ionization fraction.  Therefore, the similarities between IRIS SJI 1400\mbox{\AA} and MXIS \ion{He}{1} mostly likely indicates multi-temperature plasma co-spatial in the plane of the sky.  Meanwhile, the equilibrium ionization fraction peak for \ion{Mg}{2} directly overlaps neutral helium; although, neutral helium can exist more readily at higher temperatures (up to $\log T \approx 4.7$). 

In agreement with the statistics provided by \cite{antolin2015}, the physical width of individual rain blobs observed in the IRIS SJI 1400\mbox{\AA}  and 2796\mbox{\AA}  channels range from approximately $0\farcs75$ to $1\farcs5$.  As again evidenced by the close spatial coherence in Figure~\ref{fig:time_series}, the physical widths of rain blobs observed in \ion{He}{1} 10830\mbox{\AA} appear consistent with those of IRIS; however, we purposefully do not derive statistics for the \ion{He}{1} rain widths since the resolution limit fluctuates due to seeing variability.  That said, all evidence points to the \ion{He}{1} emission being tightly coupled with the other diagnostics which suggests either a high degree of neutral ion coupling within the magnetized rain or possibly rapid ionization of any neutral helium that slips outside of the local rain environment.

\begin{figure*}[htb!]
\centering
\includegraphics[width=0.975\textwidth]{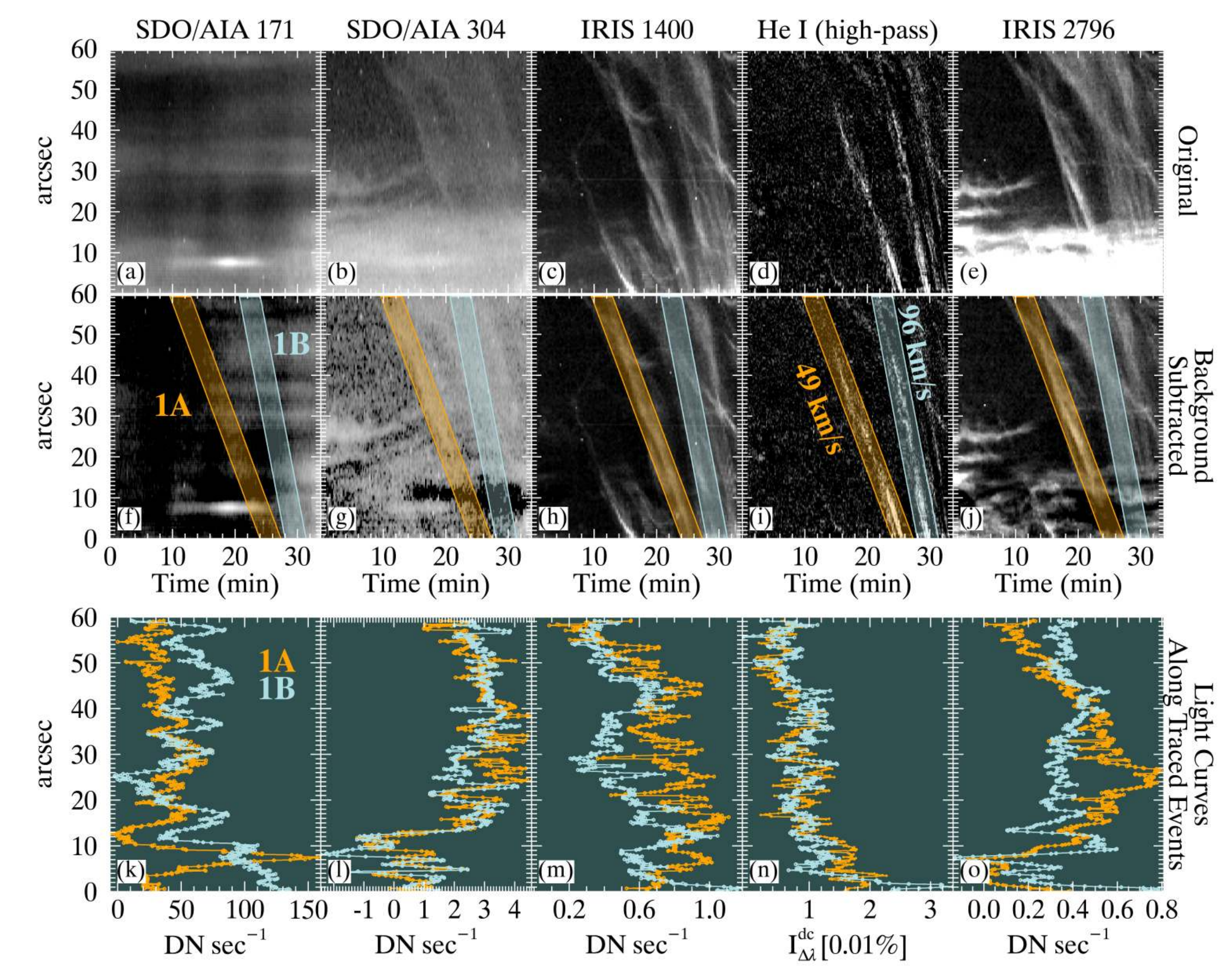}
\caption{Top: Space-time diagrams of all spectral channels extracted along the solid orange line given in Figure~\ref{fig:all_data_aligned}c.  Distance along the loop is measured in arcsec from the terminus nearest the solar limb.  Middle: The same diagrams with an approximation for the background emission subtracted and two separate rain events identified as `1A' and `1B', respectively exhibiting apparent downflow velocities of 49 and 96 km s$^{-1}$. Bottom:  Spatio-temporal light curves along paths `1A' and `1B' for each spectral channel.  The reference time for the space-time diagrams is 16:38:22 UT.}
\label{fig:time_slices}
\end{figure*}

\begin{figure*}
\centering
\includegraphics[width=0.475\textwidth]{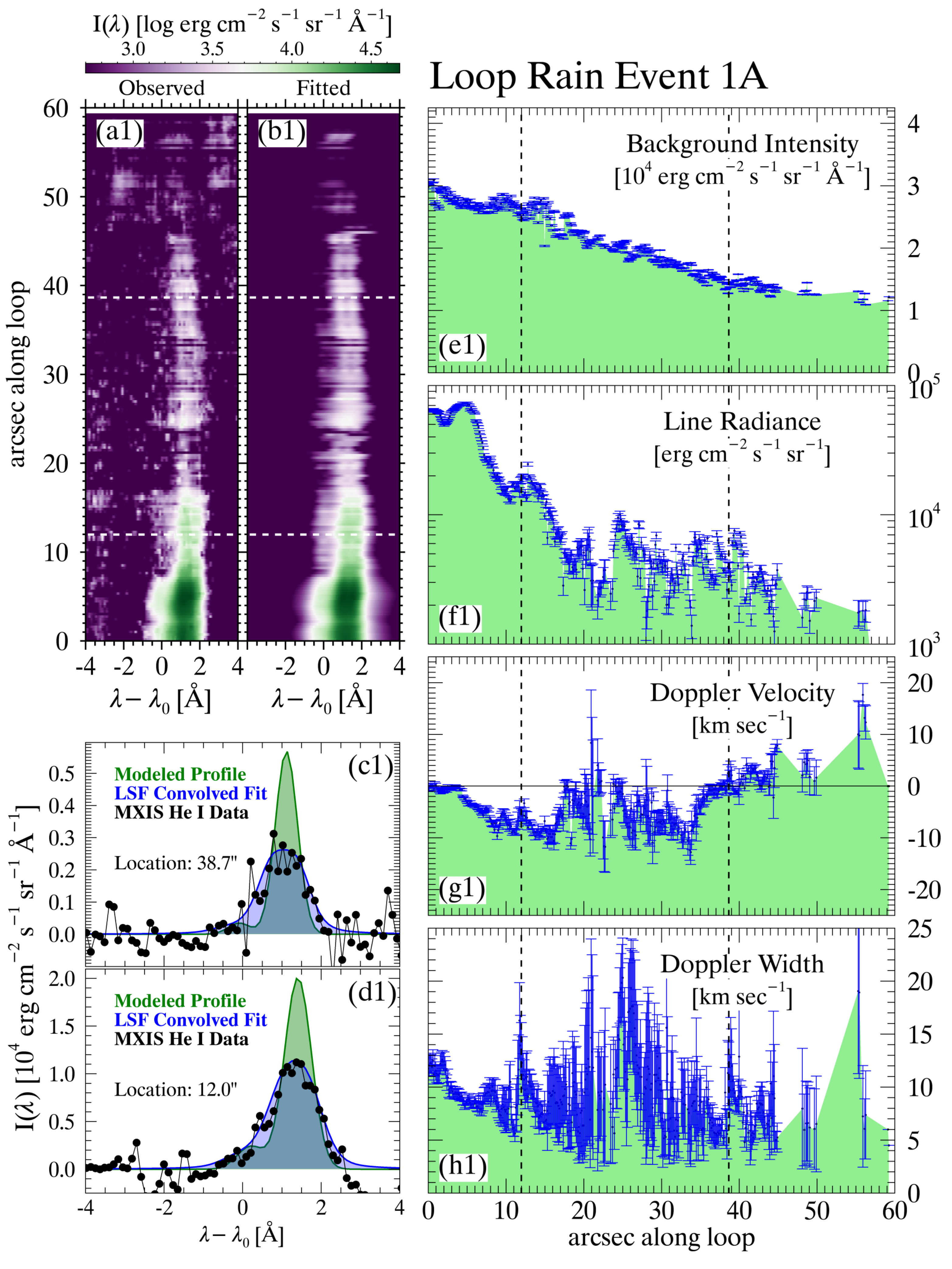}
\includegraphics[width=0.475\textwidth]{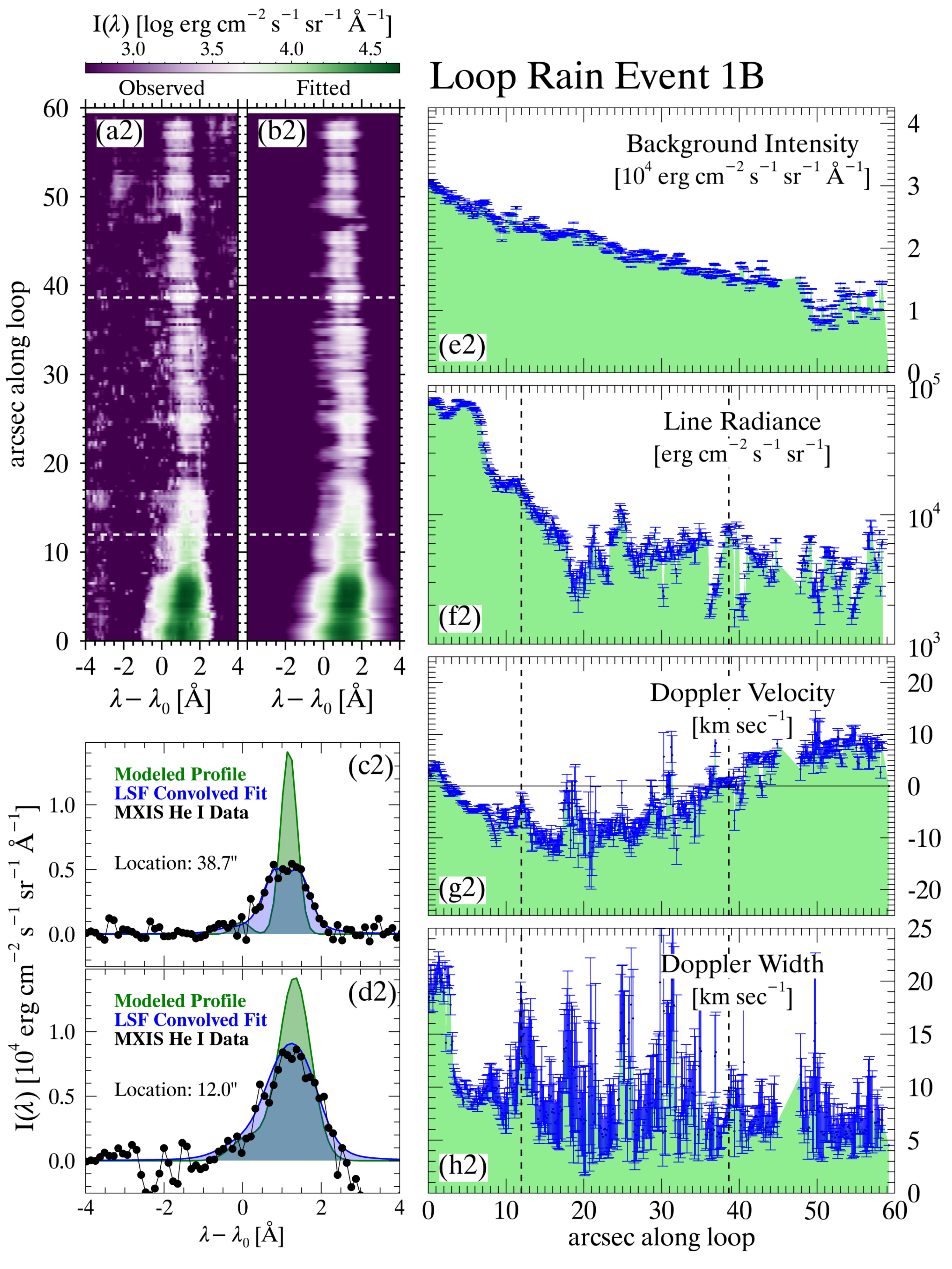}
\caption{MXIS \ion{He}{1} 10830\mbox{\AA} profiles observed in coronal rain (a1 and a2) extracted along rain paths `1A' and `1B' in Figure~\ref{fig:time_slices}.  Best fit modeled profiles are shown along each path (b1 and b2) and for the individual locations marked (c1,d1,c2,d2).  The variation of the fitted parameter as a function of distance along the paths are shown in (e1-h1,e2-h2).  Distance along the loop is measured from the terminus of the loop closest to the solar limb, and Doppler redshifts in wavelength are attributed negative Doppler velocities.}
\label{fig:rain_spectra}
\end{figure*}

\subsection{Kinematics}\label{sec:time_slices}

The neutral helium emission of coronal rain also appears dynamically coupled, or at least dynamically comparable, with the ionized species as shown by the time slice diagrams presented in Figure~\ref{fig:time_slices}.  The time slices are extracted along the center of the channel identified by dashed and solid outlines in the top row of Figure~\ref{fig:time_series}.  Prior to extraction, the data is convolved with a square $0\farcs5$ wide averaging kernel, and therefore the extracted quantities refer to averages over $0\farcs5$ scales.  The time slices are also placed onto a common temporal axis using nearest neighbor sampling.  Note that, as before, the MXIS \ion{He}{1} data presented here are sourced from the narrowband filter imaging channel and has the high-pass filter applied.  For this reason, emission from the active region prominence is not visible in the bottom portion of panel (d).  A rough approximation is made for the background structure for each time slice by averaging over the time steps just prior to the appearance of the coronal rain.  The resulting diagrams with the background subtracted are in the middle row of Figure~\ref{fig:time_slices}.  

The 34 minute duration time slices, in particular those extracted from the comparatively higher resolution IRIS and MXIS data, reveal multiple episodes of coronal rain with apparent motion along the selected trajectory.  The onset of the rain's appearance in these channels also correspond with the onset of enhanced \ion{He}{2} 304\mbox{\AA} emission observed by AIA; however, the 304\mbox{\AA} channel has much coarser resolution and the fine-scale structuring of the multiple raining episodes is only weakly present.  In the AIA 171\mbox{\AA} channel, we see relatively weak evidence for extinction of the background EUV emission, which is common in coronal rain observations of hot EUV lines due to the photoionization of neutrals \citep{gilbert2011}.  

The apparent downflow speed of individual coronal rain blobs can be well described here by a single velocity.  Two such downflows are identified in Figure~\ref{fig:time_slices} as events 1A and 1B with respective characteristic apparent speeds of 49 km s$^{-1}$ and 96 km s$^{-1}$.  We extract the intensity for each channel along both paths and display the light curves in the bottow row of Figure~\ref{fig:time_slices}.  Recognizing the limitations imposed by the background emission, we see limited evidence in this subsample of events for spatio-temporal correlations in the brightness of individual rain blobs among the various spectral channels.  The IRIS 1400\mbox{\AA} and MXIS \ion{He}{1} channels show a weak systematic increase in brightness as the blobs fall.  The other channels exhibit less systematic behavior; though event 1A shows a sizable increase and then decrease in its IRIS 2796\mbox{\AA} brightness.  It is, of course, important to note here that these time slices do not follow the full evolutionary history of the material as it first forms outside of the MXIS field-of-view. 

\subsection{\ion{He}{1} spectral analysis}\label{sec:spec_fits}

MXIS's rapid imaging spectroscopy capability allows observation and analysis of the \ion{He}{1} 10830\mbox{\AA} spectral signatures within coronal rain.  Figure~\ref{fig:rain_spectra} presents a spectral analysis of the radiometrically calibrated profiles extracted for events 1A and 1B identified in Figure~\ref{fig:time_slices}.  As before, the spectral data is first convolved with a square $0\farcs5$ wide averaging kernel, and therefore the extracted quantities refer to averages over $0\farcs5$ scales.  The observations are shown in panels (a1) and (a2) for the two respective events, and example spectral profiles are shown along with model fits (discussed below) in panels (c1), (d1), (c2), and (d2).  The data itself readily indicates the presence of the \ion{He}{1} triplet along the majority of event 1A and all of event 1B; though, residual instrumental artifacts are also present.

For each spectral profile, we fit an LSF-convolved model spectral profile using least squares minimization.  The model consists of the addition of two Gaussian profiles and a scalar value for the background (scattered light) intensity.  The two Gaussian profiles represent (1) the blue component of the \ion{He}{1} triplet with rest wavelength at 10829.0911\mbox{\AA} and (2) the two blended transitions of the \ion{He}{1} red component near 10830.295\mbox{\AA}.  The free parameters of the model include the background intensity, the peak intensity of the red spectral component, the ratio of the blue to red component peak intensity, the Doppler line width, and Doppler shift.  Errors in the fitted parameters are estimated using repeated fits of modeled profiles with different realizations of the measured noise.

Considering the spectral data results from integration times of only 105 msec, it is encouraging that the signal to noise of the spectra allows full analysis.  Satisfactory model fits are achieved for the majority of the observed profiles; however, we eliminate from further consideration all profiles for which the maximum of the convolved model profile is less than three times the root sum square of the fit residuals.  Figure~\ref{fig:rain_spectra} panels (c1), (d1), (c2), and (d2) demonstrate modeled profile fits with the background subtracted for select spectra along the two rain events.  The shapes of the modeled profile prior to and after LSF-convolution are both shown.  In part, these plots emphasize that the total line radiance is more reliably recovered in these measurements than the peak line brightness due to the influence of the line spread function.  

For events 1A and 1B, panels (e1) and (e2) of Figure~\ref{fig:rain_spectra} show the radial dependence of the background scattered light contribution.  The total line radiance shown in panels (f1) and (f2) peaks near $8\times10^{4}$ erg cm$^{-2}$ s$^{-1}$ sr$^{-1}$ having increased for locations closer to the solar limb in agreement with the trends observed in the narrowband filter data shown in Figure~\ref{fig:time_slices}.  The inferred ratio of the blue to red component intensity is not shown in the figure but is generally small (median value of 0.14) but is also subject to errors as large as 0.25.  The relatively weak signature of the blue component is visible in the modeled profiles show in panels (c1), (d1), (c2), and (d2).  The Doppler velocities along each loop, but especially 1B, show blueshifts of $\approx-10$ km s$^{-1}$ at the top of the traced loop that gradually become redshifts as large as $\approx 10$km s$^{-1}$ (for locations between $10\arcsec$ and $30\arcsec$ along the loop).  Near the lower portion of the loop, the Doppler shifts decline.  Such behavior, given the rather steady apparent motion in the plane of the sky (see Figure~\ref{fig:time_slices}), suggests geometrical changes of the observed loop relative to the line-of-sight.  For event 1B, a $10$ km s$^{-1}$ change in the Doppler shift, assuming a constant $96$ km s$^{-1}$ translational velocities, implies only an $(\arctan 10/96 \approx)$  $6^{\circ}$ change in the loop inclination.

The median Doppler line width for events 1A and 1B is $\approx$8.5$\pm$2 km s$^{-1}$ which corresponds to a plasma temperature of $\approx$ 17,000 K ($\log T \approx 4.23$) under the assumption of no unresolved motions.  At the lower end of event 1B ($\lesssim 4\arcsec$ from the near limb terminus) the line widths increase to greater than $16$ km s$^{-1}$, which are non-thermal on account of the formation temperatures of \ion{He}{1}.  This could be the influence of unresolved plasma components in the vicinity of the active region prominence (see Figure~\ref{fig:all_data_aligned}).  

%%%%%%%%%%%%%%%%%%%%%%%%%%%%%%%%%%%%%%%%%%%%%%%%%%%%%%%%%%%%%%%%%%%
%%%%%%%%%%%%%%%%%%%%%%%%%%%%%%%%%%%%%%%%%%%%%%%%%%%%%%%%%%%%%%%%%%%
%%%%%%%%%%%%%%%%%%%%%%%%%%%%%%%%%%%%%%%%%%%%%%%%%%%%%%%%%%%%%%%%%%%

\begin{figure*}[htb!]
\centering
\includegraphics[width=0.975\textwidth]{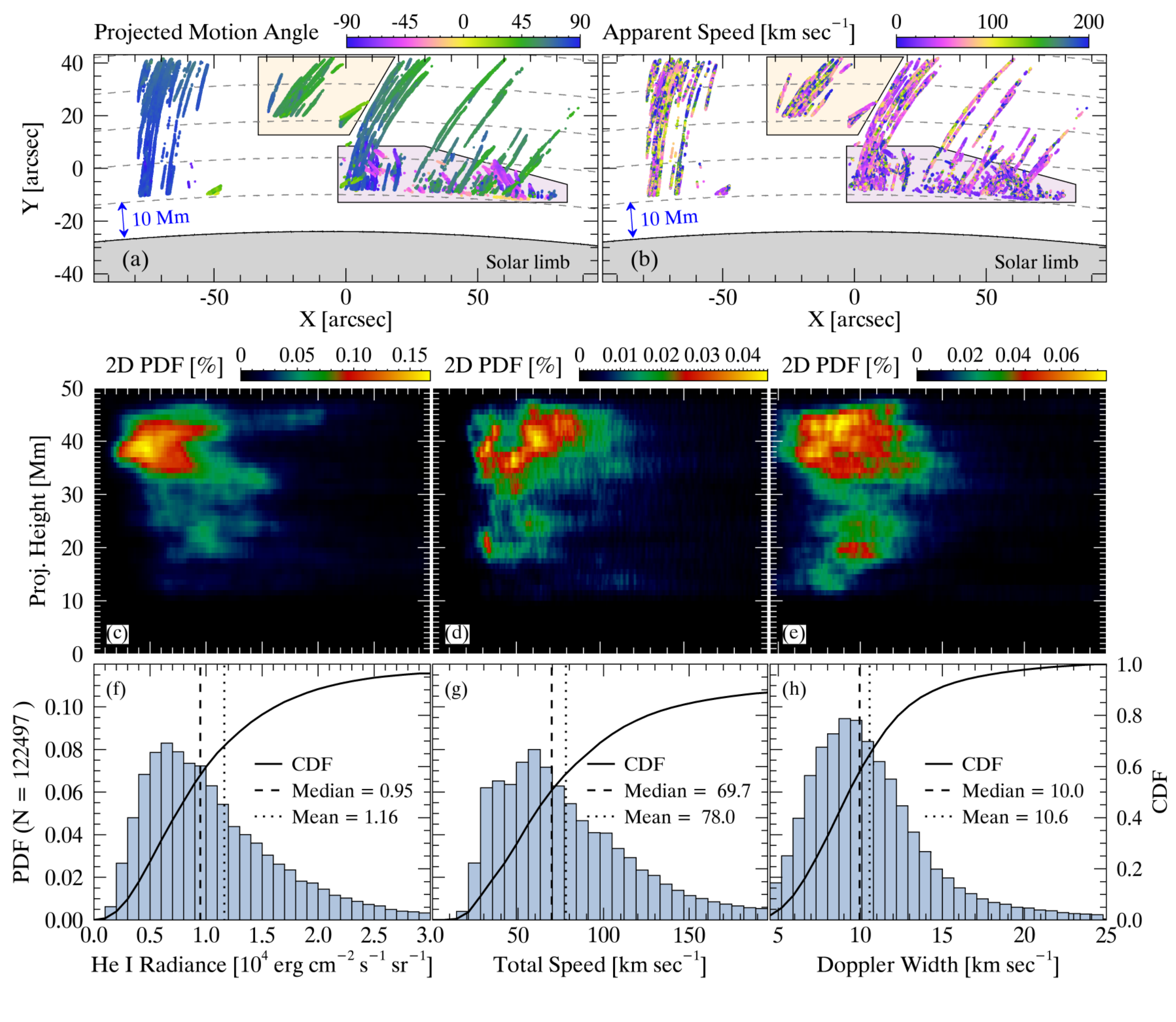}
\caption{Statistical properties of \ion{He}{1} coronal rain identified using the Multi-Dimensional Rolling Hough Transform method.  The $1.4 \times 10^{5}$ data points corresponding to coronal rain overlap in a projection onto the plane of the sky.  (a) The projected angle of apparent motion in the plane of the sky.  The two polygons indicate regions of downflowing surge material and material that overlaps with the lower lying prominence. These regions are not included in the statistical analysis below. (b) The apparent velocity of the material along its trajectory as projected in the plane of the sky, $v_{\parallel}$. (c-e) Two-dimensional probability distribution functions (PDFs) for the projected height above the limb relative to \ion{He}{1} line radiance, total speed ($\sqrt{v_{Dopp}^{2} + v_{\parallel}^{2}}$), and Doppler line width (in velocity units).  (f-h) 1D PDFs and cumulative distribution functions (CDF) of each variable along with median and mean values.}
\label{fig:rht_results}
\end{figure*}

\section{Statistical properties of \ion{He}{1} Coronal Rain}\label{sec:statistics}

Expanding the above analysis, we now derive the statistical properties of all \ion{He}{1} coronal rain produced and observed in our targeted region.  We take advantage of the Multi-Dimensional Rolling Hough Transform (MD-RHT) developed by \cite{schad2017_rht} to detect in an automated fashion coronal rain features in a time-series imaging data set.  In the cited work, the MD-RHT technique has already been applied with success to the IRIS SJI 1400\mbox{\AA}  observations described here.  

\subsection{Application of the Multi-Dimensional Rolling Hough Transform}

The Multi-Dimensional Rolling Hough Transform (MD-RHT) consists of a computer vision technique that automates tasks traditionally accomplished manually via time slice analysis and consequentially is well suited for the detection and kinematic analysis of coronal rain.  We apply this technique on the high-pass filtered MXIS \ion{He}{1} narrowband imaging data by first interpolating the data set to a uniform temporal cadence ($\delta t = 8.5$ sec), as required.  The interpolation conservatively uses nearest neighbor sampling in time.  For feature segmentation in the spatial domain, we create binary versions of the data using appropriate thresholds after it has been first smoothed temporally with a 168 second wide kernel.  In the temporal domain, segmentation is accomplished using the same zero-phase-lag bidirectional difference filter as in \cite{schad2017_rht} with a difference width of $\pm4$ time steps ($\pm33.6$ sec).   We restrict the application to areas in the field-of-view 10 Mm above the solar limb so as to disregard spicular motions, and we further disregard a large portion of the prominence body.

The results of the MD-RHT process are shown in Figure~\ref{fig:rht_results}.  Only data points for which the MD-RHT provides statistically significant identifications (see \cite{schad2017_rht}) are included.  Furthermore, we only consider data points about which $>75\%$ of its neighboring pixels within an $0\farcs5 \times 0\farcs5$ area also have good MD-RHT results.  This ensures that the identified features have a sufficient width before we extract the spectrum averaged over an $0\farcs5 \times 0\farcs5$ area for model fitting.  These filters result in $\approx 3 \times 10^{5}$ data points corresponding to material with apparent motion.  Eliminating the regions corresponding to the solar surge downflow and where there is line-of-sight confusion near the active region prominences, there are $1.4\times10^{5}$ data points corresponding to coronal rain. 

Apparent translational velocities along each feature in the plane of the sky ($v_{\parallel}$) are derived as part of the MD-RHT and shown for all data points in panel (b) of Figure~\ref{fig:rht_results}.  Errors in $v_{\parallel}$ scale with the apparent speed as shown in Figure 10 of \cite{schad2017_rht}.  For each data point corresponding to coronal rain, we extract the \ion{He}{1} profile from the MXIS spectrograph data and perform the same model fitting as in Section~\ref{sec:spec_fits}.  $85\%$ of the profiles are successfully fit.  In panels (c), (d), and (e) of Figure~\ref{fig:rht_results}, two-dimensional probability distributions function (PDFs) are shown for the projected height of the coronal rain as functions of the total line radiance, total material speed ($\sqrt{v_{Dopp}^{2} + v_{\parallel}^{2}}$), and the Doppler line width.  Below these panels are 1D PDFs of each fitted spectral parameter. 

\subsection{Average \ion{He}{1} properties}

Changes in the total line radiance and/or Doppler width as a function of height can indicate the material evolves as it falls and/or becomes radiatively excited in different ways.  \cite{antolin2015}, for example, discovered decreases in the average height of material observed in progressively cooler spectral bands, spanning chromospheric to coronal temperatures, which was interpreted to be a signature of runaway cooling ongoing as the material falls.  Furthermore, spectral lines radiatively excited by non-flat portions of the solar spectrum, for example, resonance lines like \ion{He}{1} 584\mbox{\AA} and 537\mbox{\AA}, are subject to Doppler dimming and brightening effects; however, these effects are expected to be very small for the \ion{He}{1} triplet lines \citep{labrosse2007a,labrosse2007b}. 

The 2D-PDFs in Figure~\ref{fig:rht_results} do not show evidence that the average \ion{He}{1} 10830\mbox{\AA} properties are changing over the restricted range of heights studied here.  This very well may be a consequence of the rain being fairly mature in its formation before entering the MXIS field-or-view, or its possible the loops do not have coherent evolution across the observed region.  Either way, for heights up to 50 Mm, and especially between 30 and 50 Mm where the majority of the rain is detected, the properties are fairly uniform albeit with broad distributions in the 1D PDFs shown on the bottom of the figure.  We note that the shape of the 2D-PDF for total speed of the \ion{He}{1} rain is different from that of IRIS 1400\mbox{\AA} presented in \cite{schad2017_rht} primarily because the selection of data points differs greatly.  The earlier study includes the solar surge and covers a slightly different time period.

Regarding average properties, we learn from this data that approximately half of the observed material has \ion{He}{1} 10830\mbox{\AA} line radiances greater than $10^{4}$ erg cm$^{-2}$ s$^{-1}$ sr$^{-1}$.  The total speeds are dominated by the apparent translation velocities with a median of $\sim70$ km s$^{-1}$.  The distribution of Doppler line widths has a median value of 10 km s$^{-1}$ and is largely consistent with thermal widths in absence of unresolved motions; however, no constraint for potential non-thermal line broadening is available here.

%%%%%%%%%%%%%%%%%%%%%%%%%%%%%%%%%%%%%%%%%%%%%%%%%%%%%%%%%%%%%%%%%%%
%%%%%%%%%%%%%%%%%%%%%%%%%%%%%%%%%%%%%%%%%%%%%%%%%%%%%%%%%%%%%%%%%%%
%%%%%%%%%%%%%%%%%%%%%%%%%%%%%%%%%%%%%%%%%%%%%%%%%%%%%%%%%%%%%%%%%%%

\section{Error Estimates for \ion{He}{1} Triplet Spectropolarimetric Diagnostics of Coronal Rain}

The observed constraints for \ion{He}{1} triplet coronal rain emission allow us to assess the sensitivity of future spectropolarimetric diagnostics of coronal rain that include the magnetic field vector.   Coronal rain, as shown, occurs on small scales and evolves quickly; therefore, the most important consideration is what role photon noise has on the sensitivity at dynamically-limited integration times.  We use forward modeling of the \ion{He}{1} polarization and a Monte Carlo inversion experiment that take into account realistic photon-dominated measurement errors.  Other technical complications are ignored.   Since observations of different multiplets differently influence and/or reduce errors when observed alone or together \citep{casini2009}, we consider synthetic observations of both of the two brightest orthohelium multiplets, \textit{i.e.} 10830\mbox{\AA} and 5876\mbox{\AA} (D$_{3})$.  In addition to the normal reduction of errors provided by multiple measurements with uncorrelated noise, the increased diversity of the individual multiplet's polarizability provides additional physical constraints for inverted solutions. 

\begin{figure}[htb!]
\centering
\includegraphics[width=0.475\textwidth]{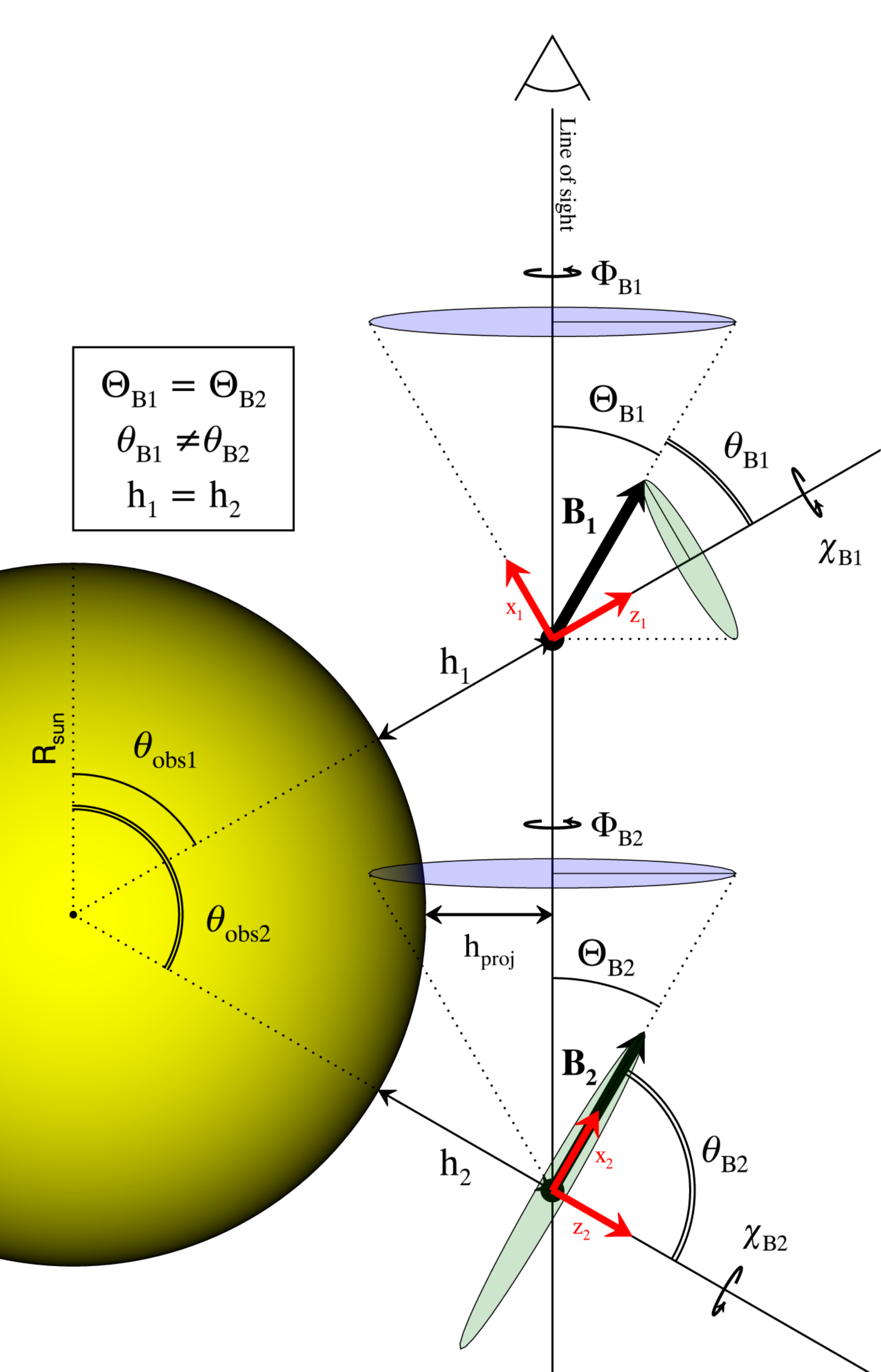}
\caption{Geometry of scattering event modeled for the \ion{He}{1} system by \textsc{Hazel} ($\chi_{obs} = 0$; $\gamma_{obs}=0$).  Two magnetic field vectors ${\bf B_{1}}$ and ${\bf B_{1}}$ illustrate field vectors with identical inclination $\Theta_{B}$ and azimuth $\Phi_{B}$ angles relative to the line-of-sight but with different inclination angles $\theta_{B}$ relative to the local solar vertical due to different values of the observation angle $\theta_{obs}$.  The two vectors, as described in the text, are distinguishable in the saturated Hanle regime and thus provide constraints on the scattering angle as an inverted parameter.}
\label{fig:scatter_geometry}
\end{figure}

\subsection{HAZEL Forward Modeling} 

Forward calculations performed here use the \textsc{Hazel} (an acronym for HAnle and ZEeman Light) computer program \citep{asensio_ramos2008}.  \textsc{Hazel} rigorously solves the statistical equilibrium equations for the atomic density matrix of the \ion{He}{1} triplet system under the multiterm atom formalism \citep[see][]{landi_2004}.  This self-consistently includes atomic level polarization induced by radiative anisotropies, the modification of quantum coherences by the Hanle effect, and the Zeeman and Paschen-Back effects.  It also includes coherences between different J-levels for the same term, which are important for magnetic strengths for which different J-levels experience crossings and repulsions.

We consider magnetic field strengths in the range of 0 G to 100 G consistent with coronal loops hosting raining material $\gtrsim$ 10 Mm above strong active regions.  This range encompasses the Hanle sensitive regime for both multiplets and the level crossings between 10 G and 100 G for D$_{3}$.  It represents the most limiting case for error estimation---the linearly polarized component of the Zeeman effect becomes measurable for B $\gtrsim$ 350 G.  While the D$_{3}$ multiplet is less bright than 10830\mbox{\AA}, its transitions have larger critical field strengths $B_{crit}^{up}$ for the operation of the Hanle effect than 10830\mbox{\AA} \citep{asensio_ramos2008}.  As a result, the D$_{3}$ Hanle sensitive regime extends up to $\approx$ 70 G, while 10830\mbox{\AA} is fully saturated for fields greater than $\approx 8$ G, making D$_{3}$ a valuable constraint for weak coronal rain magnetic fields. 

All synthesized profiles are calculated in the optically thin limit assuming plasma with constant emissive properties localized to a single point in space.  The optically thin assumption is justified by the weak blue component observed in the \ion{He}{1} 10830\mbox{\AA} rain spectra, which is roughly consistent with the small blue-to-red brightness ratio ($\sim$0.12) for optically thin plasmas \citep{centeno2008}.  The \textsc{Hazel} modeling approach does not specify the mechanism by which the \ion{He}{1} triplet levels are populated.  As described by \cite{centeno2008} and \cite{asensio_ramos2008}, for conditions ranging solar parameters, the number of bound-bound transitions for the triplet dominate the number of photoionizations from the triplet level, thereby concluding that the triplet system atomic density matrix elements are governed by the radiative transitions within the triplet system itself without significant influence from singlet helium.

\ion{He}{1} model parameters included in \textsc{Hazel} are the magnetic field intensity $B$, its inclination $\theta_{B}$ with respect to the reference axis (i.e. $\hat{z}$, the solar vertical) and its azimuth $\phi_{B}$ about $\hat{z}$, the Doppler (thermal) line width $v_{T}$, Doppler velocity $v_{mac}$, and a line damping parameter $a$.  It is further constrained by the height of the material above the solar surface $h$, which in turn governs the anisotropy factor of the presumed cylindrically symmetric pumping photospheric radiation field.  Two angles, $\chi_{obs}$ and $\theta_{obs}$, shown in Figure 1 of \cite{asensio_ramos2008}\footnote{The subscript 'obs' have been added here for clarification}, designate the observational geometry by specifying the direction of the line-of-sight with respect to the reference axes, while the angle $\gamma_{obs}$ denotes the reference direction for the positive Stokes Q.  In this work, $\chi_{obs}$ and $\gamma_{obs}$ are both taken to be 0. The choice of $\chi_{obs}=0$ establishes the reference direction for $\phi_{B}$ as the plane containing the solar vertical and the line-of-sight.    The resulting geometry is schematically illustrated in Figure~\ref{fig:scatter_geometry} (discussed further below) showing two different magnetic field vectors \textbf{\textit{B}}$_{1}$ and  \textbf{\textit{B}}$_{2}$.  $\Theta_{B}$ and $\Phi_{B}$, respectively, refer to the inclination and azimuth of the magnetic field vector with respect to the line-of-sight direction.

\subsection{Inverted Parameters and Degeneracies} 

Typically \ion{He}{1} spectropolarimetric inversions fit the model line parameters ($B$, $\theta_{B}$, $\phi_{B}$, $v_{T}$, $v_{mac}$, $a$)  using a fixed observational geometry;  $\theta_{obs}$ and $h$ are held constant with $\theta_{obs}$ usually constrained by the projected location of the material on the solar disk or in the plane of the sky.  However, for material at large heights and/or projected above the solar limb, $\theta_{obs}$ and the height $h$ are not well determined by observations; only the projected height of the material $h_{proj}$ may be directly measured off limb.  For low-lying material near the limb, \textit{e.g.} active region prominences, it may suffice to assume the material lies near the plane of the sky ($85^{\circ} \lesssim \theta_{obs} \lesssim 95^{\circ}$), as done in \cite{casini2009}. However, this is not the case for coronal rain that may form at heights up to and greater than $\sim100$ Mm.  Here we show that with a known projected height $h_{proj}$, the scattering angle $\theta_{obs}$ and the true material height $h$ may be directly inverted along with the \ion{He}{1} line parameters, thereby determining the raining material's location in 3D space. 

Figure~\ref{fig:scatter_geometry} helps to illustrate how the observational geometry influences the polarized signatures for the \ion{He}{1} triplet and gives insight into how this allows it to be fit via inversion when the projected height is known.  The magnetic field vectors \textbf{\textit{B}}$_{1}$ and  \textbf{\textit{B}}$_{2}$ have identical inclination $\Theta_{B}$ and azimuth $\Phi_{B}$ angles relative to the line-of-sight and the same height $h$, but each vector denotes material located on opposite sides of the Sun relative to the observer.  For spectral lines polarized only via the Zeeman effect, two such vectors are indistinguishable on account of the polarization amplitude and direction being only sensitive to the orientation of the magnetic field vector relative to the line-of-sight.  In contrast, the two vectors are distinguishable for the \ion{He}{1} triplet because the atomic level polarization and the Hanle effect introduce an additional dependency of the polarized amplitude and direction on the orientation of the magnetic field relative to the local solar vertical.  

To quantify the dependency of the polarization on the field orientation relative to both the line-of-sight and the local solar vertical directions, in part to understand what limits exist on constraining the scattering angle via inversion, it is useful to introduce simplified expressions for the frequency-dependent emission coefficients along the line-of-sight direction $(\vec{\Omega}_{0})$.  Here we present the more simplified two level atom approximation in the saturated regime of the Hanle effect (assuming weak field Zeeman effect), which offers a suitable approximation to the more complicated behavior of the multiterm \ion{He}{1} system.  The following have been derived for magnetic-dipole transitions by \cite{casini1999} and \cite{landi_2004}; and the transformation for electric-dipole (E1) transitions like the \ion{He}{1} triplet is straightforward.  Adopting notation from \cite{landi_2004} (Section 13.5) and using the reduced statistical tensor representation for the atomic density matrix, we can write for E1 transitions: 
\begin{widetext}
\begin{subequations} 
\label{eqn:emiss}
\begin{align} 
\epsilon_{i}(\nu,\vec{\Omega}_{0})_{E1} &= C_{J_{u}J_{l}} \left [ 1 +  \frac{\omega_{J_{u}J_{l}}^{(2)} }{4\sqrt{2}}(3\cos^{2}\Theta_{B}-1)(3\cos^{2}\theta_{B}-1)[\sigma_{0}^{2}(\alpha_{u}J_{u})]_{v}   \right ] \hat{\phi}(\nu)  \label{eqn:ei}  \\
\epsilon_{q}(\nu,\vec{\Omega}_{0})_{E1} &=  \frac{3 C_{J_{u}J_{l}} \omega_{J_{u}J_{l}}^{(2)} [\sigma_{0}^{2}(\alpha_{u}J_{u})]_{v} }{4\sqrt{2}} \left [(1-3\cos^{2}\theta_{B})\sin^{2}\Theta_{B}\cos 2 \Phi_{B} \right ]\hat{\phi} (\nu) \label{eqn:eq}    \\ 
\epsilon_{u}(\nu,\vec{\Omega}_{0})_{E1} &=  \frac{3 C_{J_{u}J_{l}} \omega_{J_{u}J_{l}}^{(2)} [\sigma_{0}^{2}(\alpha_{u}J_{u})]_{v} }{4\sqrt{2}} \left [(1-3\cos^{2}\theta_{B})\sin^{2}\Theta_{B}\sin 2 \Phi_{B} \right ]\hat{\phi} (\nu)  \label{eqn:eu}  \\ 
\epsilon_{v}(\nu,\vec{\Omega}_{0})_{E1} &=  C_{J_{u}J_{l}} \cos\Theta_{B} \nu_{L} \left [ \bar{g} + \frac{\Delta}{2}(3\cos^{2}\theta_{B} -1) [\sigma_{0}^{2}(\alpha_{u}J_{u})]_{v}  \right ] \frac{\partial \hat{\phi} (\nu) }{\partial \nu} \label{eqn:ev}  
\end{align} 
\end{subequations}
\end{widetext} 
where $[\sigma_{0}^{2}(\alpha_{u}J_{u})]_{v}$ is the ``reduced'' fractional atomic alignment in the vertical frame aligned with the solar vertical and
\begin{equation}
C_{J_{u}J_{l}} = \frac{h\nu}{4\pi}\mathcal{N}\sqrt{2J_{u}+1}A(\alpha_{u}J_{u}\rightarrow\alpha_{l}J_{l})\rho_{0}^{0}(\alpha_{u}J_{u}).
\end{equation}
$\omega_{J_{u}J_{l}}^{(2)}$ is a scalar coefficient depending on the atomic parameters, $\nu_{L}$ is the Larmor frequency (dependent on the field intensity $B$), $\bar{g}$ is the effective Land\'e factor, $\Delta$ is a factor that depends on atomic parameters and the Land\'e factors of the levels, and $\hat{\phi} (\nu)$ is the frequency dependence of the line profile shape.  These expressions encode the geometrical dependencies of the polarized line emission on $\theta_{B}$, $\Theta_{B}$, and $\Phi_{B}$, which are related via the scattering angle and the spherical law of cosines by 
\begin{equation} \label{eqn:cos_trans}
\cos \theta_{B} = \cos \theta_{obs} \cos \Theta_{B} - \sin \theta_{obs} \sin\Theta_{B}\cos\Phi_{B}.
\end{equation}
As before, the reference direction of the linear polarization is the plane containing the solar vertical and the line of sight.  Note that equations 20-22 of \cite{asensio_ramos2008} are consistent with Equations~\ref{eqn:eq} and~\ref{eqn:eu}.  Finally, the projected height and true height are related by
\begin{equation}
h_{proj} = \cos \left ( \frac{\pi}{2} - \theta_{obs} \right ) (R_{sun} + h) - R_{sun}.
\end{equation}

Returning to the illustration of Figure~\ref{fig:scatter_geometry}, we observe that the inclination $\theta_{B1}$ (relative to the vertical) of \textbf{\textit{B}}$_{1}$ is substantially greater than that of \textbf{\textit{B}}$_{2}$.  As drawn, $\theta_{B1}=30^{\circ}$ and $\theta_{B2}=90^{\circ}$.  Due to the $(1-3\cos^{2}\theta_{B})$ dependence of Equations~\ref{eqn:eq} and~\ref{eqn:eu}, this implies that  \textbf{\textit{B}}$_{1}$ and \textbf{\textit{B}}$_{2}$ will have opposite signs (directions) of linear polarization.  This is known as the Van Vleck effect with the ``magic" Van Vleck angle being $54.74^{\circ}$.  It is this  $\theta_{B}$ dependence of $\epsilon_{q}$ and $\epsilon_{u}$ that discriminates  \textbf{\textit{B}}$_{1}$ from  \textbf{\textit{B}}$_{2}$.  It is worth pointing out that $\epsilon_{v}$ also has a weak dependence on $(1-3\cos^{2}\theta_{B})$, meaning that the Stokes V amplitude is slightly different for  \textbf{\textit{B}}$_{1}$ and  \textbf{\textit{B}}$_{2}$ despite having the same longitudinal magnetic field strength.  

We now ask what conditions give the same (or approximately the same) values for the Stokes vectors to discuss degeneracies and/or limited constraints present in Equations~\ref{eqn:emiss} when the projected height is known.  In principle, the combination of the angular dependencies and the height dependence of  $[\sigma_{0}^{2}]_{v}$ limits the presence of true degeneracies; however, the height dependence of $[\sigma_{0}^{2}]_{v}$ is weak, and noise in any measurement may make it unable to distinguish between similar solutions.  Also, note that $\epsilon_{q}$ and $\epsilon_{u}$ have no dependence on the field intensity, and $\epsilon_{v}$ (in the limit of weak $[\sigma_{0}^{2}]_{v}$) constrains only the longitudinal component of the field intensity ($B\cos\Theta_{B}$).  However, this does not inherently limit our ability to infer the true field intensity B.  Any scaling of the field intensity used to match an observed Stokes V amplitude requires a change in $\Theta_{B}$, and, as seen in Equation~\ref{eqn:cos_trans}, there is no way to preserve $\theta_{B}$, which sets the linear polarization amplitude, under a perturbation in $\Theta_{B}$ by using the scattering angle $\theta_{obs}$ to compensate.

One approximately degenerate solution is immediately apparent given a transformation of $\Phi_{B}$ to $\Phi_{B}^{\prime} = \Phi_{B}+\pi$ when $\Theta_{B}^{\prime} = \Theta_{B}$ is held fixed.  $\epsilon_{q}$ and $\epsilon_{u}$ are preserved if the scattering angle switches from $\theta$ to $\theta^{\prime} = \pi - \theta$, which implies $\theta_{B}^{\prime} = \pi -  \theta_{B}$.  This ``180-degree ambiguity'' is an \textit{approximate} degeneracy since the change in $\theta_{B}$ does have a weak effect on $\epsilon_{v}$ that would be difficult to observe due to the presence of measurement noise. 

Remaining potential ambiguities may be determined by solving for the combinations of $\{\theta_{obs}, \Phi_{B}\}$ that give equivalent values for $\epsilon_{q}$ and $\epsilon_{u}$ assuming fixed values of the field intensity in the saturated Hanle regime, constant $\Theta_{B}$, and neglecting the height dependence of $[\sigma_{0}^{2}]_{v}$.  $\epsilon_{q}$ and $\epsilon_{u}$ then have two terms that can vary to compensate the sign of each other.  The potential ambiguities are given for $\Phi_{B}^{\prime} = \{\Phi_{B},\Phi_{B}+\pi,\Phi_{B}-\pi/2,\Phi_{B}+\pi/2\}$.  For the first two potential $\Phi_{B}^{\prime}$ values, we can solve for $\theta^{\prime}$ subject to the constraint $(1-3\cos^{2}\theta_{B}^{\prime}) =(1-3\cos^{2}\theta_{B})$, while for the later two potential $\Phi_{B}^{\prime}$ values, the constraint is $-(1-3\cos^{2}\theta_{B}^{\prime})=(1-3\cos^{2}\theta_{B})$. Using Equation~\ref{eqn:cos_trans} and tangent half-angle substitution, the solutions are given by the roots of the quadratic equation
\begin{equation} \label{eqn:ambig}
at^{2} + bt + c = 0 
\end{equation}
where
\begin{eqnarray}
a &=& (\mathcal{K} - \cos\Theta_{B}'),  \nonumber \\
b &=& (-2\sin\Theta_{B}^{\prime}\cos\Phi_{B}^{\prime}), \nonumber\\
c &=& (\mathcal{K} + \cos\Theta_{B}'), \nonumber \\
\mathcal{K}  &=& \begin{cases}
\mp \cos\theta_{B} & \text{ if } \Phi_{B}^{\prime} = \{\Phi_{B},\Phi_{B}+\pi\}, \\ 
\mp\sqrt{\frac{2}{3}-\cos^{2}\theta_{B}}& \text{ if } \Phi_{B}^{\prime} = \{\Phi_{B}\pm\pi/2\},
\end{cases} \nonumber \\
t &=& \tan\frac{\theta_{obs}^{\prime}}{2} \nonumber.
\end{eqnarray}
In some cases, the degenerate solutions give values of $\theta_{obs}$ that may imply an unphysical, large height of the observed material (especially in the case of coronal rain).  Here we restrict the solutions to $60^{\circ} < \theta_{obs} < 120^{\circ}$.  Furthermore, no solutions exist for the $\Phi_{B} \pm \pi/2$ degenerate angles when $\theta_{B} < 35^{\circ}$ or $\theta_{B} >145^{\circ}$; however, there may still be local minima caused by this potential degeneracy. 

\begin{figure*}[htb!]
\centering
\includegraphics[width=0.975\textwidth]{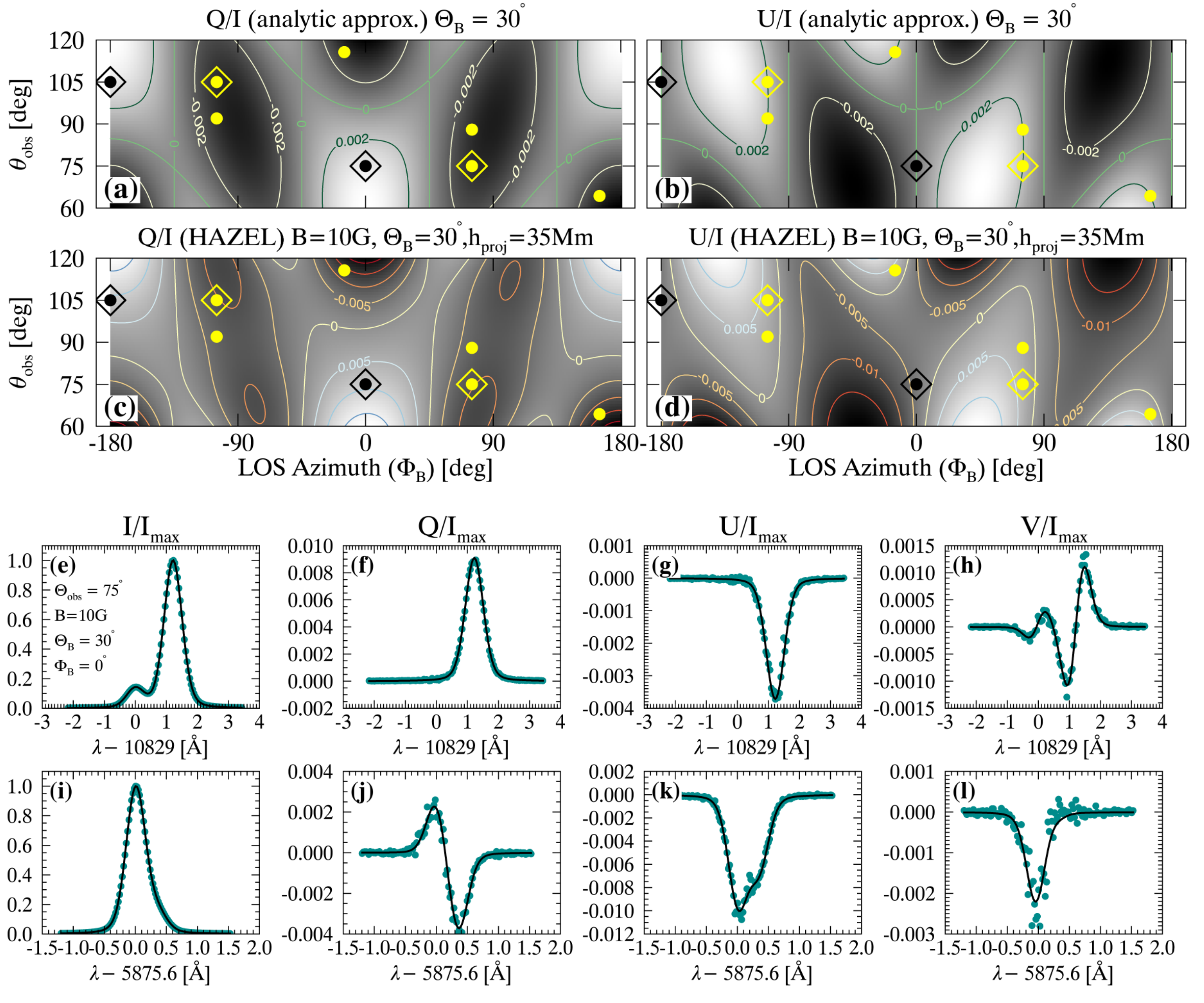}
\caption{Identification of candidate degenerate solutions for inversions that include fits for the magnetic field vector and the scattering angle $\theta_{obs}$ with fixed projected height above the solar limb.  Panels (a) and (b) give the geometric dependencies of the analytic approximate value for Q and U (Equations~\ref{eqn:eq} and~\ref{eqn:eu}) for the $J_{u}=2 \rightarrow J_{l}=1$ transition at 10830.3398\mbox{\AA} with a constant value of 0.0525 used for $\sigma_{0}^{2}|_{v}$ assumed and a constant line-of-sight inclination of $\Theta_{B}=30^{\circ}$.  Panels (c) and (d) display the same information as (a) and (b) but use the full \textsc{Hazel} forward model of the multiplet including height dependence of $\sigma_{0}^{2}|_{v}$.  Black and yellow filled circles show locations of candidate degenerate solutions for two sample vectors according to Equation~\ref{eqn:ambig}; though, in the absence of noise only the ``180-degree ambiguity'' solutions marked with diamonds generally apply.  (e-i) give example fits to synthetic observations of the 10830\mbox{\AA} and D$_{3}$ multiplets with the noise expected for a 4 meter aperture with 10\% transmission sampling scales of $(0\farcs5)^{2}$ for an integration time of 5.5 seconds.}
\label{fig:inv_noise}
\end{figure*}

Figure~\ref{fig:inv_noise} gives a graphical representation of the candidate degenerate solutions for two magnetic field vectors.  The geometric dependencies of Q and U under the approximate solution (\textit{i.e.} the angular dependent terms of Equations~\ref{eqn:eq} and~\ref{eqn:eu}) and as calculated by \textsc{Hazel} are shown.  Using the global optimization approach described in the next section, we find that given expected noise amplitudes for a large aperture solar telescope, typically only the ``180-degree ambiguity" applies. 

\subsection{Inversion Approach}

Given synthetic Stokes profiles with known parameters and with noise applied, we numerically optimize the set of \emph{inverted} parameters that provide the best reduced chi-squared fit (\textit{i.e.} Equation 23 of \cite{asensio_ramos2008}) to the synthetic profiles under test.  The free parameters include all of the \textsc{Hazel} \ion{He}{1} optically thin line parameters ($B$, $\theta_{B}$, $\phi_{B}$, $v_{T}$, $v_{mac}$) except the line damping parameter $a$, which is held fixed throughout.  In addition, the fitted parameters include the total line radiance and the scattering angle $\theta_{obs}$, but the projected height is assumed to be known.

While \textsc{Hazel} has a built-in inversion module for singly-observed \ion{He}{1} multiplets, here we use only its forward modeling engine and instead use optimization modules from the SciPy Python library \citep{jones2001} including the Differential Evolution (DE) algorithm \citep{storn1997} and the wrapper for the Levenberg-Marquardt (LM) method implemented in MINPACK \citep{more1978}.  This allows simultaneous inversions of the 10830\mbox{\AA} and D$_{3}$ multiplets and the ability to fit the scattering angle constrained by a constant projected height.  DE is a metaheuristic optimization routine with features and performance similar to the \textsc{Pikaia} algorithm \citep{charbonneau1995}, which has previously been used for solar inversion problems.  Unlike LM, it evolves stochastically without knowledge of the gradient of the objective function.  It can locate degenerate solutions using repeated applications, as done using \textsc{Pikaia} in \cite{schad2016}, and used here to test for potential ambiguities (see Figure~\ref{fig:inv_noise}). Below, we use a minimal number of DE iterations to provide the initialization for the LM fit so that our results are not dependent on initial guesses. 

\subsection{Monte Carlo Error Simulation}

Our Monte Carlo simulation estimates errors for inverted coronal rain parameters by generating and inverting a database of synthetic observables with realistic measurement noise applied.  Multiple scenarios are studied, including those consistent with currently existing 1.5 meter class facilities, \textit{e.g.} the 150 cm GREGOR solar telescope \citep{schmidt2012} and the 160 cm Goode Solar Telescope \citep{goode2010}, and the upcoming 4 meter aperture Daniel K. Inouye Solar Telescope \citep[DKIST:][]{rimmele2015}.  The database includes 4000 models (2000 for B$<$10 G and 2000 for  B$>$10 G) selected via Latin hypercube sampling (LHS) of parameters defined in Table~\ref{tbl:model_para}, and thus our results represent an average over the respective domains.  For the \ion{He}{1} 10830\mbox{\AA} line radiance and Doppler width, median values derived from our observations for angular sizes of $0.5''$ are used.  The \ion{He}{1} D$_{3}$ line radiance is set by a fixed energy ratio E$_{10830}$/E$_{5876}=6$ ($\approx$11 in units of photons) consistent with the optically thin prominence models of \cite{labrosse2004}. 

The additive photon noise applied to the modeled profiles assumes uncorrelated normally distributed values with a standard deviation of $\sqrt{I(\lambda)}$ for Stokes I and $\sqrt{I(\lambda)}/\xi$ for the polarized spectra.  $\xi$ denotes polarimetric modulation efficiency \citep{deltoro2000} and is set to $\xi=1/\sqrt{3}$ for all polarized states.  The maximum integration time of 5.5 seconds is motivated by the angular sample ($0\farcs5 \approx 360$ km) crossing time for material with median translational velocities of $\sim 70$ km s$^{-1}$.  The measured signal amplitude depends on the line brightness and the effective aperture of the telescope, which include the transmission of post-focus instrumentation.  Before considering specific instrumentation, we model separate cases of 1.5 and 4 meter telescopes with effective apertures set by the respective collecting area and a total transmission of 10\%.  

The inversions sequentially invoke the DE and LM algorithms.  Four generations of the differential evolution algorithm are first carried out with specific bounds enforced on the candidate solution, including a field intensity within $\pm 20$ G of the known model parameters and all angles within $\pm 12.5^{\circ}$ of the modeled values.  This step provides a stochastically sourced initial guess for the LM-based inversion step.  The parameter bounds are only used in the DE step in order to limit the parameter space searched to that where the known global minimum exists.  The LM step proceeds until convergence criteria are met.  The inverted profiles must achieve a reduced chi-squared of less than 1.05 to be considered a valid solution.  The results are discussed in Section~\ref{sec:error_results}.

\begin{deluxetable}{lccl}
\tablecaption{Parameters for Monte Carlo Error Model \label{tbl:model_para}}
\tablecolumns{4}
\tabletypesize{\scriptsize}
\tablenum{1}
\tablehead{
\colhead{Parameter} & 
\colhead{Modeled Range} & 
\colhead{Inverted} & 
\colhead{Units}
}
\startdata
\multicolumn{4}{l}{\textbf{\ion{He}{1} Line Parameters}}  \\ 
\hline
\ion{He}{1} 10830\mbox{\AA} Radiance                &    $10^{4}$                 &  \checkmark   & erg cm$^{-2}$ s$^{-1}$ sr$^{-1}$    \\
Optical Depth                                       &      Thin 	                      &  \xmark            &  \nodata  \\
Doppler Width ($v_{T}$)                                     &       10                         &  \checkmark  & km s$^{-1}$       \\
Doppler Velocity ($v_{mac}$)                                 &        0                         & \checkmark   &  km s$^{-1}$     \\
B field strength\tablenotemark{a} ($B$)     &     0 --100                 &  \checkmark  &  gauss     \\
B field inclination ($\theta_{B}$)                         &     0 -- 180                 &  \checkmark  & degrees \\
B field azimuth ($\chi_{B}$)                                     &     -180 -- 180           &  \checkmark  &  degrees \\
Projected Height ($h_{proj}$)                                 &     20 -- 50                  &  \xmark            &   Mm  \\
Scattering Angle ($\theta_{obs}$)                                 &    60 -- 120                 &  \checkmark  &  degrees \\
Line damping  ($a$)                                     &     0.2                          &   \xmark          & dimensionless \\
E$_{10830}$/E$_{5876}$                  &    6                              &  \checkmark  & \nodata \\
\hline
\multicolumn{4}{l}{\textbf{ Observation Parameters}}  \\ 
\hline
Spatial sample		     &    $0\farcs5 \times 0\farcs5$                      &  &  arcsec$^{2}$ \\ 
Integration time           &    5.5                                                &  & sec \\ 
Spectral resolution\tablenotemark{b}  ($R$)                       &   125000          &    & \nodata  \\
Eff. Aperture - (400 cm) &  $0.1(\pi (200)^{2}$)   &   & cm$^{2}$       \\ 
Eff. Aperture - (150 cm) &  $0.1(\pi (75)^{2}$)   &   & cm$^{2}$       \\
\hline
\multicolumn{4}{l}{\textbf{DKIST Instrumentation Parameters\tablenotemark{c}}}  \\ 
\hline
VISP spatial sample\tablenotemark{d}    & $0\farcs213 \times 0\farcs5$  & & arcsec$^{2}$ \\
DL-NIRSP spatial sample\tablenotemark{d}		     &    $0\farcs5 \times 0\farcs5$       &  &  arcsec$^{2}$ \\ 
Integration time          &    5.5                                                 &  & sec \\ 
VISP spectral res. (R)                       &   65000          &    & \nodata  \\
DL-NIRSP spectral res. (R)             &    65000          &    & \nodata  \\
VISP Eff. Aperture  &  $0.012(\pi (200)^{2}$)   &   & cm$^{2}$       \\ 
DL-NIRSP Eff. Aperture &  $0.041(\pi (200)^{2}$)   &   & cm$^{2}$       \\
\enddata
\tablenotetext{a}{Sub-ranges of 0-10 G and 10-100 G are considered separately.}
\tablenotetext{b}{Spectral resolution ($R$) is given as $\lambda/(2\Delta\lambda)$, where $\Delta\lambda$ is the spectral sample width of the synthetic profiles.}
\tablenotetext{c}{Instrument parameters selected based on design-phase performance estimates.  See text for details.}
\tablenotetext{d}{Spatial sampling here includes spatial averaging of detector pixels.}
\end{deluxetable}

\subsection{Estimates for DKIST first-light instrumentation}

First light instrumentation under fabrication for DKIST has been designed to support a broad scientific mission \citep{elmore2014}.  Using performance predictions for the Visible Spectropolarimeter \citep[VISP:][]{dewijn2012} and the Diffraction-Limited Near-Infrared Spectropolarimeter (DL-NIRSP), we further refine our DKIST error predictions for coronal rain observations by performing the same simulation described above but with parameters representative of particular instruments.  Here, we assume the use of VISP for observations of D$_{3}$ and DL-NIRSP for 10830\mbox{\AA}.  

Using the provided instrument performance calculator, we expect VISP to reach an end-to-end transmission of $\sim1.2\%$ at the D$_{3}$ multiplet.  Meanwhile, the spatial sample for VISP, which is a single slit scanning spectropolarimeter capable of simultaneous observations in three spectral windows, is controlled by its slit width and spatial sampling along the slit.  The widest available slit ($0\farcs213$) is assumed to be used here so as to achieve the highest observable flux for the rain observations, and we further assume co-addition of $0\farcs5$ along the slit.  The spectral resolution is $R\sim$65,000.  

DL-NIRSP is a fiber-optic based integral field spectropolarimeter with three different spatial sampling modes of varying instantaneous fields-of-view.  Prioritizing field coverage, we assume the use of the wide-field mode with an effective spatial sample size of $\sim0\farcs5$ and an instantaneous field of view of $\sim 28\arcsec \times 19\arcsec$.  Some uncertainty exists in the total end-to-end transmission of DL-NIRSP using the wide-field fiber, but here it is estimated to be $4.1\%$ for observations at 10830\mbox{\AA}.   The spectral resolution is expected to be similar to the wide-slit VISP case. 

\begin{deluxetable*}{l|DDD|DDD}
\tablecaption{Inverted Parameter Confidence Intervals Defined as the Width of the Error Distribution Containing 68\%[95\%] of the Inverted Models \label{tbl:error_estimates}}
\tabletypesize{\small}
\tablecolumns{7}
\decimals
\tablenum{2}
\tablehead{\colhead{ } & \multicolumn{6}{c}{0 G $<$ B $<$ 10 G} & \multicolumn{6}{c}{10 G $<$ B $<$ 100 G} }
\startdata
             &  \multicolumn{2}{c}{\bf{D$_{3}$}}  &  \multicolumn{2}{c}{\bf{10830}} &  \multicolumn{2}{c}{\bf{D$_{3}$+10830}} &   \multicolumn{2}{c}{\bf{D$_{3}$}} &  \multicolumn{2}{c}{\bf{10830}} &  \multicolumn{2}{c}{\bf{D$_{3}$+10830}} \\
\hline
                                         & \multicolumn{12}{c}{150 cm Aperture (10\% Transmission)}  \\
\hline
$B$ (G)                                    &  1.6[11]  &  2.7[14]  &  0.5[5]  &  9.9[32]  &  7.1[44]  &  2.6[14]   \\
$\theta_{B}$ ($^{\circ}$)                &  5.6[28]  &  2.7[17]  &  0.6[3]  &  2.0[7]  &  0.7[9]  &  0.4[1]   \\
$\chi_{B}$ ($^{\circ}$)                  & 12.5[57]  & 10.5[44]  &  2.4[19]  &  5.8[21]  &  2.6[24]  &  1.4[7]   \\
$\theta_{obs}$ ($^{\circ}$)              &  3.9[20]  &  3.9[26]  &  0.9[4]  &  4.1[14]  &  1.9[18]  &  1.0[5]   \\
\hline
                                          & \multicolumn{12}{c}{400 cm Aperture (10\% Transmission)}  \\
\hline
$B$ (G)                                    &  0.6[3]  &  1.2[11]  &  0.2[4]  &  3.6[15]  &  3.0[31]  &  0.9[4]   \\
$\theta_{B}$ ($^{\circ}$)                &  2.5[12]  &  1.8[14]  &  0.2[1]  &  0.8[2]  &  0.3[5]  &  0.1[0]   \\
$\chi_{B}$ ($^{\circ}$)                  &  4.7[32]  &  5.8[34]  &  0.8[9]  &  2.1[7]  &  1.4[17]  &  0.4[2]   \\
$\theta_{obs}$ ($^{\circ}$)              &  1.4[7]  &  2.2[19]  &  0.3[1]  &  1.5[5]  &  0.8[13]  &  0.3[1]   \\
\hline
                                          & \multicolumn{12}{c}{400 cm Aperture + DKIST Instrumentation\tablenotemark{a}}  \\
\hline
$B$ (G)                                    &  3.2[21]  &  2.0[17]  &  0.8[19]  & 17.9[52]  &  4.2[33]  &  1.6[7]   \\
$\theta_{B}$ ($^{\circ}$)                & 10.2[41]  &  2.1[15]  &  0.7[20]  &  3.4[12]  &  0.5[7]  &  0.3[1]   \\
$\chi_{B}$ ($^{\circ}$)                  & 21.9[101]  &  7.1[44]  &  2.9[53]  &  9.4[34]  &  1.9[22]  &  0.9[3]   \\
$\theta_{obs}$ ($^{\circ}$)              &  7.2[31]  &  3.2[23]  &  1.1[32]  &  6.9[23]  &  1.1[16]  &  0.7[2]   \\
\enddata
\tablenotetext{a}{Instrument parameters selected based on design-phase performance estimates. See text for details.}
\end{deluxetable*}

\subsection{Results of the Monte Carlo Error Model}\label{sec:error_results}

Error estimates for the three cases described above are given in Table~\ref{tbl:error_estimates}.  We report the $68\%$ ($\pm1\sigma$) and $95\%$ ($\pm2\sigma$) confidence intervals defined as the width of the error distribution containing 68\% and 95\% of the inverted models; however, note that the $95\%$ confidence interval can be influenced by a small sample of modeled field vectors with large errors.  Once again, these errors represent an average over the modeled domain.  The true errors will vary depending primarily on the field intensity and orientation.  Since the error in the total radiance, Doppler width, and Doppler velocity are insignificant here, only the magnetic field parameters and scattering angle errors are included in the table.  For each case, the results are given for inversions for each multiplet separately and when combined. 

For weaker magnetic fields (0 G $<$ B $<$ 10 G), the D$_{3}$ multiplet, despite its lower total radiance, exhibits a smaller inverted $1\sigma$ error in the magnetic field intensity compared to 10830\mbox{\AA} when the effective aperture is the same at both wavelengths.  Likely, this can be attributed to its extended range in Hanle sensitivity, which at these field strengths exhibit a significant response to the field intensity.  Meanwhile, its angular errors are greater than that of the brighter 10830\mbox{\AA} multiplet, which is expected due to its lower signal-to-noise ratio.  In contrast, at larger field strengths (10 G $<$ B $<$ 100 G), 10830\mbox{\AA} exhibits smaller error for all parameters compared to D$_{3}$.  When both multiplets are observed simultaneously, the error in all parameters are reduced further. 

In the case of DKIST with its first-light instrumentation, the different transmissions and spatial samples of VISP and DL-NIRSP have a noticeable effect on the resulting error estimates, especially in comparison to the idealized 10\% transmission examples.  However, the combination of VISP and DL-NIRSP observations of D$_{3}$ and 10830\mbox{\AA} prove to be a powerful diagnostic of coronal rain magnetism.  Based on the half-width of the 68\% confidence intervals, we predict an average $\pm1\sigma$ magnetic field error of $\pm 0.4$ G and $\pm 0.8$ G, respectively, for weak and intermediate magnetic field intensities.  The errors in the angular parameters are one to a few degrees.  At the $2\sigma$ level (\textit{i.e.} the 95\% confidence intervals), errors increase significantly for the weak fields suggesting a fraction of poorly-constrained models.  If we calculate alternatively the 90\% confidence interval width, its width is 6 G, which is narrower than the 95\% confidence interval for models with B $>$ 10 G, which is 7 G.  These values may be marginally refined using more models in the synthetic database at the cost of longer computing time; however, here the upper bound of the 2$\sigma$ error, based on the half-width of the confidence intervals, is taken to be $\pm 3.5$ G. 

%%%%%%%%%%%%%%%%%%%%%%%%%%%%%%%%%%%%%%%%%%%%%%%%%%%%%%%%%%%%%%%%%%%
%%%%%%%%%%%%%%%%%%%%%%%%%%%%%%%%%%%%%%%%%%%%%%%%%%%%%%%%%%%%%%%%%%%
%%%%%%%%%%%%%%%%%%%%%%%%%%%%%%%%%%%%%%%%%%%%%%%%%%%%%%%%%%%%%%%%%%%

\section{Discussion}\label{sec:discussion}

\subsection{\ion{He}{1} triplet rain characteristics}

The MXIS imaging observations of \ion{He}{1} 10830\mbox{\AA} coronal rain indicate that neutral helium bears resemblance both to IRIS SJI transition region diagnostics of the Si \text{IV} 1393.78\mbox{\AA} and 1402.77\mbox{\AA} spectral lines ($\log T \approx 4.8$) as well as those of \ion{Mg}{2} k at 2796.35\mbox{\AA} ($\log T \approx 4$).  Generally, we have not seen evidence suggesting neutral helium is absent at the same time that the IRIS diagnostics are present, and vice versa.  This is consistent with the multi-thermal nature of coronal rain discussed by \cite{antolin2015}.  However, the MXIS observations only cover the lower 50 Mm (in projected height) of very tall coronal structures ($\gtrsim 100$ Mm).  We do not resolve where the onset of neutral helium formation occurs in relation to the IRIS diagnostics; that it is an interesting question going further considering the height dependence of different thermal diagnostics shown in \cite{antolin2015}.  The wide distribution in the observed \ion{He}{1} line widths, without any perceivable average height dependence, may suggest the various blobs of coronal rain transversing the MXIS field-of-view have inhomogenous thermal structure; however, the line widths measured here have wide errors due to observational noise.

Unlike the IRIS diagnostics, \ion{He}{1} is a neutral species and thus feels no Lorentz force due to the presence of the magnetic field.  And yet, as already shown by \cite{antolin2012sharp}, \cite{ahn2014}, and \cite{schad2016}, cool coronal downflows observed in neutral species seem to follow the same trajectories of the ionized species, seemingly tracing out magnetic field lines, and suggestive of a high degree of ion-neutral coupling.  This seems also the case for these observations.  As pointed out in the earlier studies, the expected cross-field diffusion speeds, calculated using the formalism of \cite{gilbert2002}, is typically small in comparison to the bulk velocities of the material.  Moreover, simplified numerical models of falling partially ionized coronal rain by \cite{oliver2016} indicate that a relatively small drift speed between ions and neutral is sufficient to dynamically couple the two species.  Observational studies of the decoupling of ionic and neutral species for prominences have been studied, with differing results, by \cite{khomenko2016} and \cite{anan2017}.  A very detailed spectroscopic study would be required to understand how well coupled the various species are in coronal rain. 

\subsection{Relation to prominence models}

The formation of neutral helium emission in coronal rain has not yet been studied in specific theoretically; however, many of the complications involved are well known due to the detailed non-local thermodynamic equilibrium modeling efforts for prominence material \citep[see review by][]{labrosse2010}.    In general, the total line radiance is not only a function of the column mass density and the state variables but also the external radiative environment, the dynamics, and the composition (see \cite{gouttebroze2009} and references therein).  Multi-line observations, like those for singlet and triplet helium by \cite{ramelli2012}, provide useful constraints on these effects.  Multi-spectral helium observations of coronal rain are not yet available; here, we only have observations of the \ion{He}{1} 10830\mbox{\AA} multiplet and can only make broad comparisons. 

The cylindrical thread model of \cite{gouttebroze2009} is useful to compare to our observations.  \citeauthor{gouttebroze2009} model the emission of principal lines of helium for cylindrical threads with diameters of 1000 km, and for a range of temperature, pressure, and helium abundance.  Such sizes are comparable with the physical widths of coronal rain observed here.  Therefore, we directly compare the modeled emergent intensities for \ion{He}{1} 10830\mbox{\AA} with those we observed.  For models with a constant pressure of 0.1 dyn cm$^{-2}$, a (He/H) abundance of 0.1, and temperatures ranging between 6000 and 50,000 K, \citeauthor{gouttebroze2009} infer line radiances in the range of $0.5-2 \times 10^{4}$ erg cm$^{-2}$ s$^{-1}$ sr$^{-1}$, much in line with the values observed here.  A similar agreement is seen with the models including a radially dependent temperature for a wide range of pressure and helium abundance. 

\subsection{Coronal rain \ion{He}{1} spectropolarimetric sensitivity}

The confirmed brightness of \ion{He}{1} 10830\mbox{\AA} in coronal rain, especially of the quiescent type not associated with flares or other phenomena, has important implications for its use to study the structure of the coronal magnetic field, specifically under non-equilibrium conditions.  Only a few previous studies have demonstrated measurements of the coronal field in general, including very long integration measurements of the \ion{Fe}{13} 10747\mbox{\AA} coronal emission line by \cite{lin2004} and measurements of thermal bremsstrahlung and gyroresonant radio emission by \cite{brosius2006}.  In the case of \citeauthor{lin2004}, only the longitudinal component of the field intensity was measured, and it had an amplitude of about 4 G at projected heights of $\sim100\arcsec$ ($\sim70$ Mm) off limb.   \cite{schad2016} inferred the height dependence of the vector coronal magnetic field using the \ion{He}{1} 10830\mbox{\AA} multiplet observed in absorption on disk. In that study, the total magnetic field strength for heights of 30 to 70 Mm above an active region were in the range of $10$ to $150$ G. 

Measuring the solar coronal magnetic field more routinely using forbidden coronal emission lines formed in megakelvin plasma is an essential part of DKIST's mission \citep{rimmele2015}.  Its sizable jump in collecting area (a factor of 29 greater than the Dunn Solar Telescope) allows for a significant reduction in the integration times needed to measure the megakelvin magnetic field.  Still, as implied by \cite{penn2004}, the integration times may need to be long depending on the application.  Using the same methodology of that work, we can estimate the time required for \ion{Fe}{13} observations at DKIST under ideal background-free conditions to achieve a magnetic field sensitivity equivalent to the cool \ion{He}{1} lines estimated above.  Assuming a $1\arcsec \times 1\arcsec$ angular sample, a \ion{Fe}{13} line brightness of 40 millionths of the disk intensity (width of 2\mbox{\AA}), and a $1\sigma$ error of 0.8 G, the exposure time required is $\sim460$ seconds ($7.6$ minutes).  As evidenced by this, the presence of neutral helium in coronal rain, with brightnesses two orders of magnitude larger than the hot forbidden lines and with response to the Hanle sensitive regime, offers a huge advantage in measuring the coronal magnetic field under non-equilibrium conditions.  Of course, one limitation is that coronal rain is not ubiquitous in the corona.   

%%%%%%%%%%%%%%%%%%%%%%%%%%%%%%%%%%%%%%%%%%%%%%%%%%%%%%%%%%%%%%%%%%%
%%%%%%%%%%%%%%%%%%%%%%%%%%%%%%%%%%%%%%%%%%%%%%%%%%%%%%%%%%%%%%%%%%%
%%%%%%%%%%%%%%%%%%%%%%%%%%%%%%%%%%%%%%%%%%%%%%%%%%%%%%%%%%%%%%%%%%%

\section{Summary}\label{sec:summary}

Leveraging massively multiplexed spectroscopic techniques, this work has provided new measurements and a statistical analysis of the strength and character of the neutral helium triplet formed in coronal rain.  Using coordinated IRIS measurements, we have found that the fine-scaled nature present in the ionized species correlates well, both morphologically and dynamically, with the structure in the neutral helium observations, at least within the limits of these observations.  At present, we have only found weak correlations between the brightness of the IRIS diagnostics and the neutral helium, and this is a subject for further work.  Meanwhile, our \ion{He}{1} spectral analysis has shown distributed Doppler line widths largely consistent with the formation temperatures of the neutral helium, and the small ratio between the triplet spectral components suggest the rain is optically thin at these wavelengths. 

Importantly, we have solidified the potential of using the \ion{He}{1} triplet within coronal rain for coronal magnetic field measurements.  Our analysis of coronal rain brightnesses in \ion{He}{1} 10830\mbox{\AA} has indicated a substantial portion of material with radiances greater than $10^{4}$  erg cm$^{-2}$ s$^{-1}$ sr$^{-1}$ for angular scales of $0\farcs5$.  Large aperture spectropolarimetry should be capable of making great advances in fine-scaled coronal field measurements using these diagnostics.  The techniques that we have pursued even indicate that \ion{He}{1} triplet spectropolarimetry may constrain the location of rain blobs in 3D space.  As now coronal rain is known to be a common phenomenon in active regions, continued observations of the \ion{He}{1} offer a valuable probe of localized plasma in the otherwise diffuse solar corona.

%%%%%%%%%%%%%%%%%%%%%%%%%%%%%%%%%%%%%%%%%%%%%%%%%%%%%%%%%%%%%%%%%%%

\acknowledgments

Special thanks to Haosheng Lin and Doug Gilliam for their support and assistance in conducting these measurements and to Sarah Jaeggli for a careful reading of the manuscript.  Great appreciation is also extended to Andres Asensio Ramos for developing very user-friendly tools for the modeling of the helium triplet polarization.  The National Solar Observatory (NSO) is operated by the Association of Universities for Research in Astronomy, Inc. (AURA), under cooperative agreement with the National Science Foundation.  IRIS is a NASA small explorer mission developed and operated by LMSAL with mission operations executed at NASA Ames Research center and major contributions to downlink communications funded by ESA and the Norwegian Space Centre.  \textit{SDO} data are provided courtesy of NASA/\textit{SDO} and the AIA and HMI science teams.  This research has made use of NASA's Astrophysics Data System.

%%%%%%%%%%%%%%%%%%%%%%%%%%%%%%%%%%%%%%%%%%%%%%%%%%%%%%%%%%%%%%%%%%%

\vspace{1mm}
\facilities{Dunn(MXIS), SDO(AIA), SDO(HMI), IRIS}
\software{SSWIDL, HAZEL \citep{asensio_ramos2008}, SCIPY  \citep{jones2001}}

%%%%%%%%%%%%%%%%%%%%%%%%%%%%%%%%%%%%%%%%%%%%%%%%%%%%%%%%%%%%%%%%%%%

\bibliographystyle{aasjournal}
\bibliography{schad_manuscript_v2}

%%%%%%%%%%%%%%%%%%%%%%%%%%%%%%%%%%%%%%%%%%%%%%%%%%%%%%%%%%%%%%%%%%%

\end{document}